\documentclass[twocolumn]{aastex63}

\received{January 30, 2020}
\revised{January 30, 2020}
\accepted{\today}
\submitjournal{ApJ}

\usepackage{amsmath}	
\usepackage{amssymb}
\usepackage{graphicx}
\usepackage[caption=false]{subfig}

\shorttitle{Hot spot formation in black hole accretion disks}
\shortauthors{Ripperda et al.}

\begin{document}

\title{Magnetic Reconnection and Hot Spot Formation in Black Hole Accretion Disks}

\correspondingauthor{Bart Ripperda}
\email{bripperda@flatironinstitute.org}

\author[0000-0002-7301-3908]{Bart Ripperda}\altaffiliation{Joint Princeton/Flatiron Postdoctoral Fellow}\affiliation{Center for Computational Astrophysics, Flatiron Institute, 162 Fifth Avenue, New York, NY 10010, USA}\affiliation{Department of Astrophysical Sciences, Peyton Hall, Princeton University, Princeton, NJ 08544, USA}
\author[0000-0002-0281-2745]{Fabio Bacchini}\affiliation{Centre for mathematical Plasma Astrophysics, Department of Mathematics, Katholieke Universiteit Leuven, Celestijnenlaan 200B, B-3001 Leuven, Belgium}
\author[0000-0001-7801-0362]{Alexander A. Philippov}\affiliation{Center for Computational Astrophysics, Flatiron Institute, 162 Fifth Avenue, New York, NY 10010, USA}

\begin{abstract}
Hot spots, or plasmoids, forming due to magnetic reconnection in current sheets, are conjectured to power frequent X-ray and near-infrared flares from Sgr A$^*$, the black hole in the center of our Galaxy. It is unclear how, where, and when current sheets form in black-hole accretion disks. We perform axisymmetric general-relativistic resistive magnetohydrodynamics simulations to model reconnection and plasmoid formation in a range of accretion flows. Current sheets and plasmoids are ubiquitous features which form regardless of the initial magnetic field in the disk, the magnetization in the quasi-steady-state phase of accretion, and the spin of the black hole.
Within 10 Schwarzschild radii from the event horizon, we observe plasmoids forming, after which they can merge, grow to macroscopic scales of the order of a few Schwarzschild radii, and are ultimately advected along the jet's sheath or into the disk. Large plasmoids are energized to relativistic temperatures via reconnection and contribute to the jet's limb-brightening. We find that only hot spots forming in magnetically arrested disks can potentially explain the energetics of Sgr A$^*$ flares. The flare period is determined by the reconnection rate, which we find to be between $0.01c$ and $0.03c$ in all cases, consistent with studies of reconnection in isolated Harris-type current sheets. We quantify magnetic dissipation and non-ideal electric fields which can efficiently inject non-thermal particles. We show that explicit resistivity allows for converged numerical solutions, such that the electromagnetic energy evolution and dissipation become independent of the grid scale for the extreme resolutions considered here.
\end{abstract}


\section{Introduction} 
\label{sec:intro}

Bright, coincident episodic X-ray and near-infrared flares are detected on roughly a daily basis coming from Sgr A$^*$, the supermassive black hole in the center of our Galaxy (see e.g., \citealt{baganoff2001}; \citealt{genzel2003}; \citealt{Marrone2008}; \citealt{Meyer2008}; \citealt{neilsen2013}; \citealt{fazio2018}; \citealt{boyce2019}).
The origin of these flares is generally associated with electron acceleration in a localized flaring region not larger than a few gravitational radii rather than with a global increase in the accretion rate or jet power (\citealt{markoff2001,Dodds_Eden2009,ponti2017}).
The \cite{Gravity2018} recently reported the detection of positional changes of such near-infrared flares originating from within 10 gravitational radii of the black hole.
The observed variable emission is conjectured to arise from the motion of a compact ``hot spot'' orbiting within a dynamical time scale of the compact object.

Dissipation of magnetic energy through the process of magnetic reconnection in thin current layers is conjectured as the main mechanism producing the energetic electrons powering flares and hot spots (\citealt{Yuan_2004,Broderick2005,Goodman,younsi_2015,ball2016,Li_2017,gutierrez2019}). 
If such a reconnection layer becomes thin enough it can be liable to the tearing instability, break up, and produce chains of plasmoids (\citealt{loureiro2007}) that interact, merge, grow, and are advected with the flow.  
From local studies of current sheets we know that magnetic reconnection can accelerate electrons in plasmoids to form non-thermal energy distributions (\citealt{sironi2014}; \citealt{Guo_2014}; \citealt{Werner_2015}; \citealt{rowan2017}; \citealt{werner2018}) potentially explaining the observed flaring emission. However, the formation of such reconnection layers in accretion flows has not yet been understood due to the lack of a model that captures both the dissipative microphysics and the global dynamics of the disk and jet. It is therefore essential to study reconnection in global simulations of accretion flows.

The global dynamics of black-hole accretion flows and jet formation is typically accurately modeled with ideal (i.e.,\ infinitely conductive) general-relativistic magnetohydrodynamics (GRMHD) simulations (see e.g., \citealt{porth2019}; \citealt{white2019}). While ideal GRMHD has provided significant insight in accretion flows, dissipation occurs at the grid scale and the frozen-in condition (\citealt{Alfven1942}) is broken purely due to numerical resistivity, such that it is not a reliable framework to study reconnection and plasmoid formation. 
Current sheet formation, magnetic reconnection, and plasmoid production in accretion disks have so far however only been investigated with ideal GRMHD simulations (\citealt{ball2017,kadowaki2019,nathanail2020}). 
The plasma in accretion flows like that of Sgr A$^*$ is effectively collisionless such that reconnection occurs because of kinetic effects (\citealt{parfrey2019}) that break the frozen-in condition. For example, the divergence of the electron pressure tensor essentially plays the role of an effective resistivity (\citealt{Bessho2005}).
In general-relativistic resistive magnetohydrodynamics (GRRMHD) an explicit finite resistivity $\eta$ acts as a proxy for kinetic effects, presenting the simplest model of magnetic reconnection and plasmoid formation (\citealt{loureiro2016}) in turbulent black-hole accretion flows. For Lundquist numbers $S = v_A L / \eta \gtrsim 10^4$, where $v_A$ is the Alfv\'{e}n speed and $L$ is the typical length of the current sheet, resistive reconnection becomes plasmoid-dominated and evolves independently of the resistivity $\eta$ at a universal rate of $\sim 0.01v_A$ (\citealt{bhattacharjee2009}; \citealt{uzdensky2010}). An explicit resistivity also allows for a converged evolution of electromagnetic energy density and for studying how it is dissipated through Ohmic heating. Additionally, the GRRMHD equations inherently provide information on the evolution of non-ideal electric field which can be responsible for particle acceleration, non-existent in ideal GRMHD. 

With current state-of-the-art GRRMHD simulations it has been impossible to resolve fast small-scale reconnection dynamics accurately enough to determine whether plasmoids can form and grow in accretion flows (\citealt{vourellis2019,Tomei_2019}).
Extreme resolutions are necessary to resolve current sheets that are thin enough to be liable to the plasmoid instability, while also capturing the global accretion flow dynamics.
A combination of an implicit-explicit (IMEX) time-stepping scheme (\citealt{ripperda2019b}) to capture fast reconnection dynamics, together with adaptive mesh refinement (AMR) capabilities of the Black Hole Accretion Code ({\tt{BHAC}}, \citealt{BHAC,Olivares2019}) to accurately resolve the smallest scales in the system allows us to study resistive reconnection and plasmoid formation in black-hole accretion disks for the very first time. We employ the GRRMHD module of {\tt BHAC} to investigate whether a hot spot can form and grow nearby the event horizon. Additionally, we measure the reconnection rate and investigate whether the time and length scales on which magnetic reconnection occurs and plasmoids form in the accretion flow is in accordance with the variability of the multi-wavelength emission as observed for Sgr A$^*$.

\begin{figure}
    \centering
    \includegraphics[width=0.45\textwidth,trim= 9.5cm 1cm 8cm 2cm, clip=true]{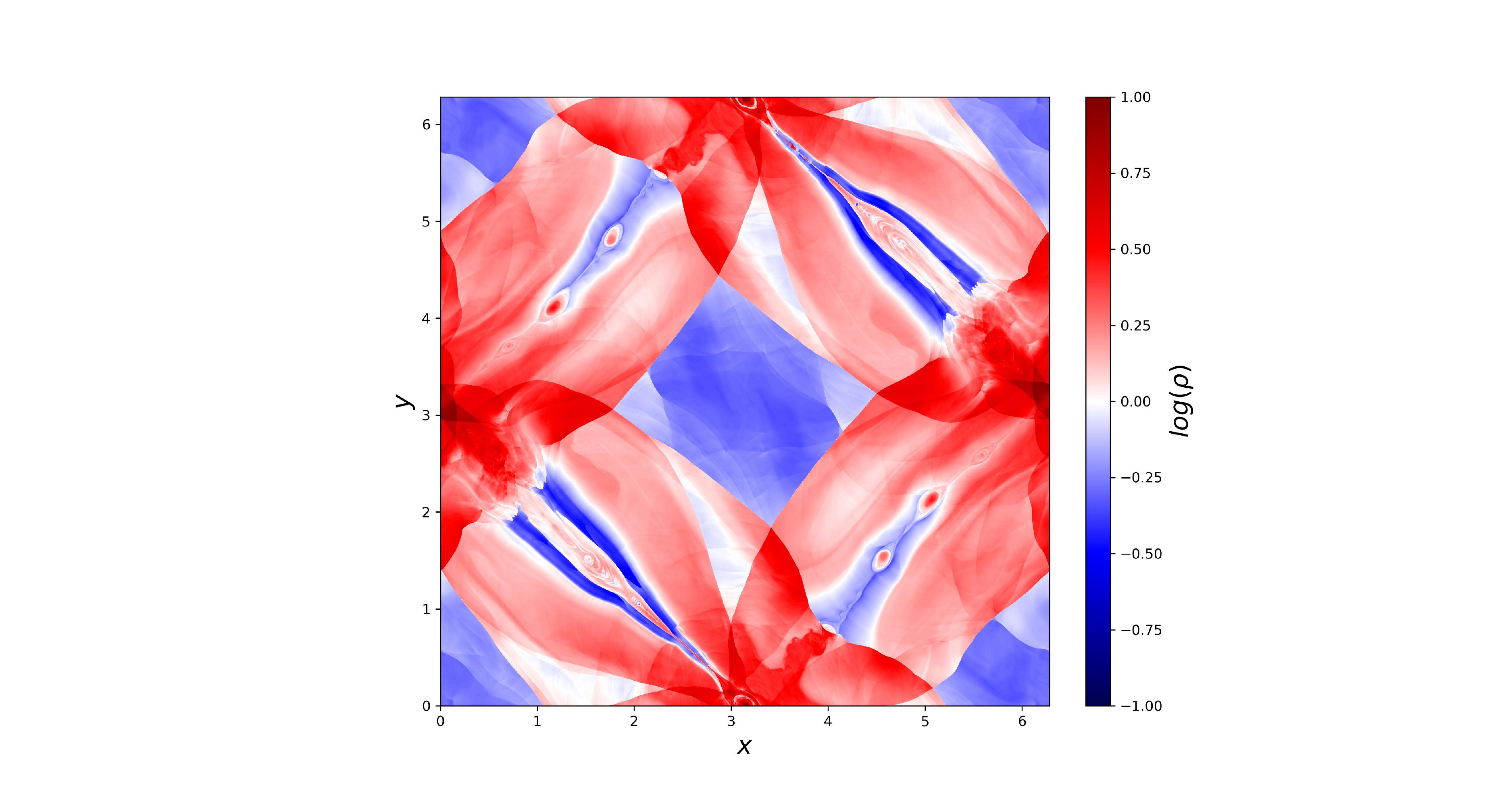}
    \caption{Rest-mass density $\rho$ for a vortex at $t=10 t_{\rm c}$ for Lundquist number $S=10^5$. The plasmoids can be recognized as overdense blobs of plasma in the thin sheets.}
    \label{fig:OTmass}
\end{figure}

\begin{figure*}
    \centering
     \includegraphics[width=0.32\textwidth,trim= 9.5cm 1cm 8cm 2cm, clip=true]{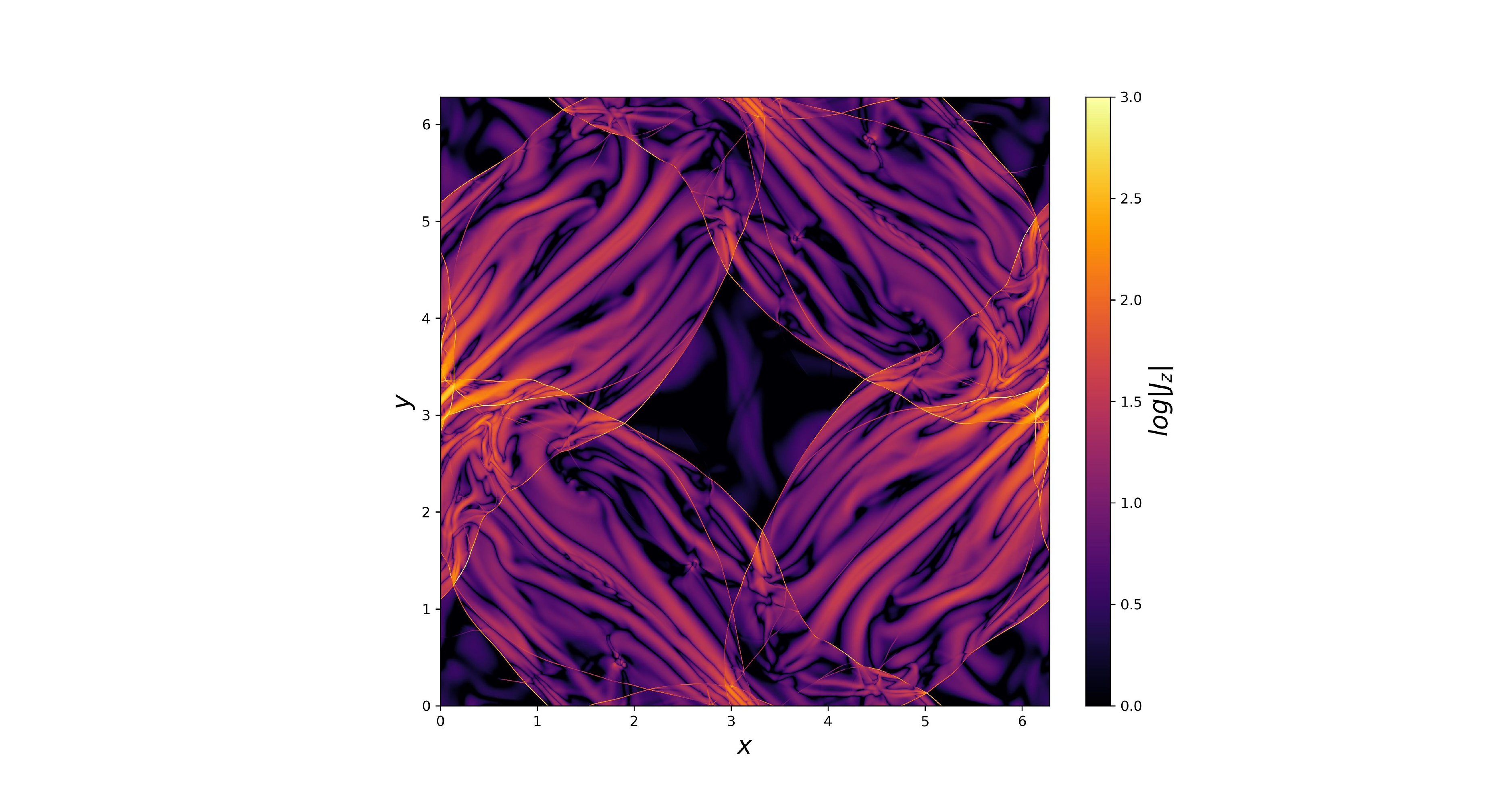}
     \includegraphics[width=0.32\textwidth,trim= 9.5cm 1cm 8cm 2cm, clip=true]{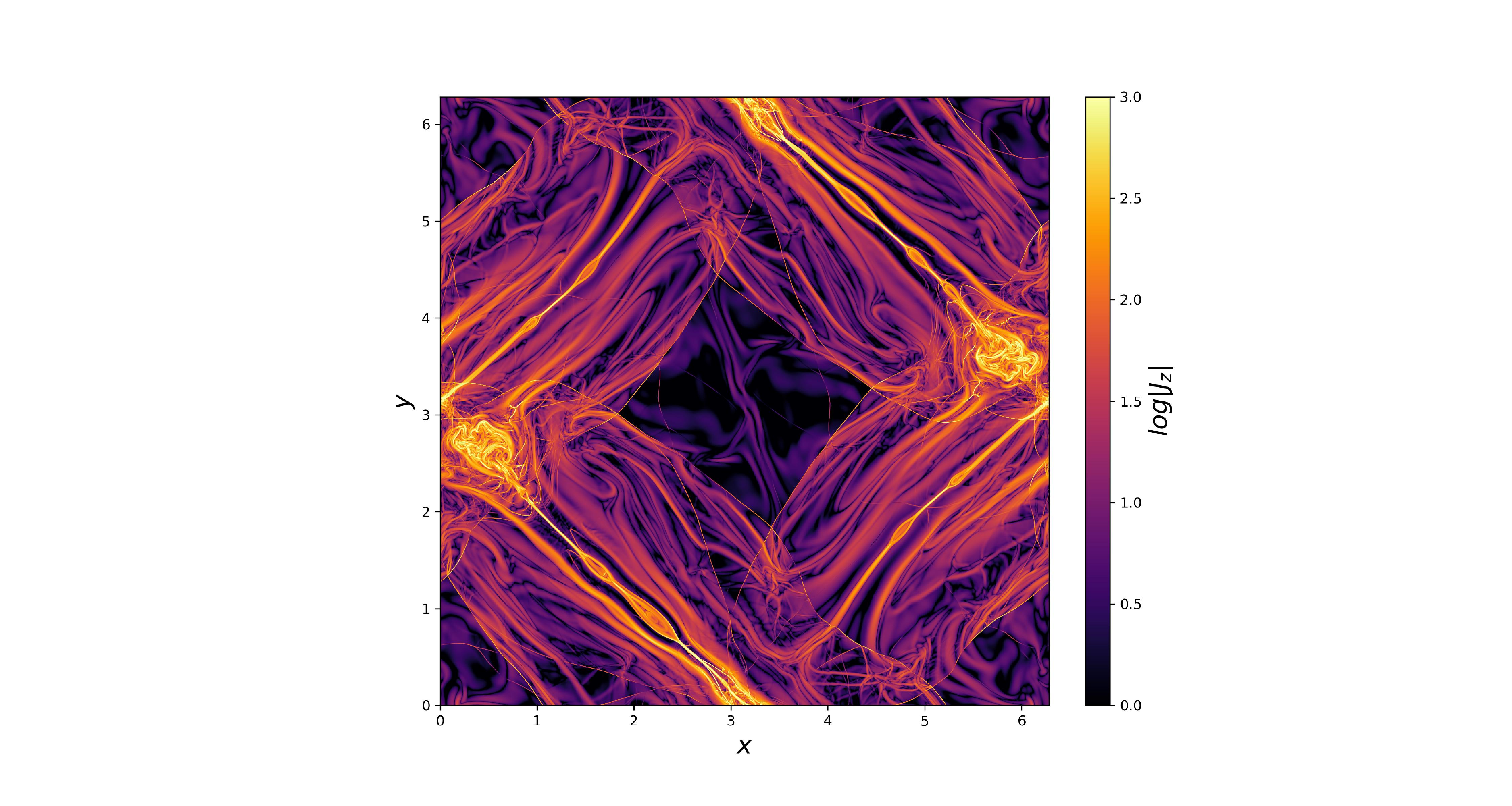}
      \includegraphics[width=0.32\textwidth,trim= 9.5cm 1cm 8cm 2cm, clip=true]{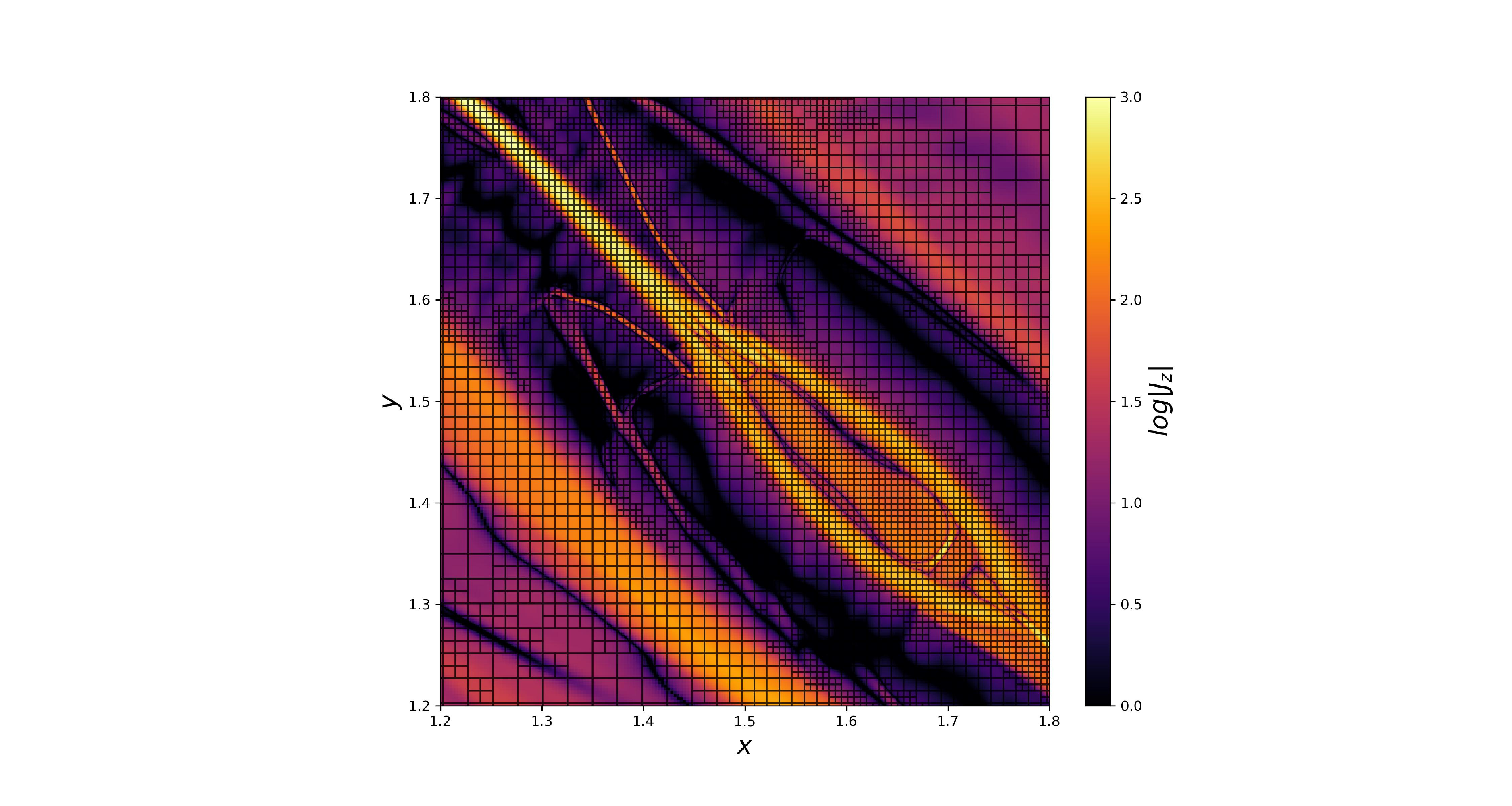}
    \caption{Logarithm of the out-of-plane current density magnitude $|J_z|$ at $t=10 t_{\rm c}$ for $S=2\times10^3$ (left), $S=2\times10^4$ (middle), and a zoom into the current sheet for $S=2\times10^4$ showing the AMR grid blocks (right), each consisting of $8\times8$ cells, in black. Both cases have an effective resolution of $8192^2$ cells in the domain. The plasmoid-unstable current sheet on the right is captured by more than ten cells over its width.}
    \label{fig:OTplasmoids}
\end{figure*}

A detailed outline of the numerical method used in this work, including a comprehensive description of the GRRMHD equations can be found in \cite{ripperda2019b}. From here onward we adopt Lorentz-Heaviside units where a factor of $1/\sqrt{4\pi}$ is absorbed into the electromagnetic fields and velocities are measured in units of the speed of light $c$. We employ a $3 + 1$ decomposition of the GRRMHD equations based on the Arnowitt-Deser-Misner formalism (\citealt{ADM}) with a $(-,+,+,+)$ signature for the spacetime metric where Roman indices run over space only i.e.,\ (1,2,3).

\section{A toy model of relativistic turbulent reconnection}

We study the properties of reconnection and formation of plasmoids in current sheets in an Orszag-Tang vortex (\citealt{orszag_tang_1979}) as a simplistic toy model for a turbulent flow. We also determine the required resolutions to capture the full process of plasmoid formation. 

\subsection{Numerical setup}
Here we assume a flat-spacetime Minkowski metric and consider a relativistic ideal gas with an adiabatic index $\hat{\gamma} = 4/3$ with an initial uniform pressure $p=10$ and rest mass density $\rho= 1$. The magnetic field $\mathbf{B} = \nabla \times \mathbf{A}$ is obtained from a vector potential $\mathbf{A} = (0,0,A_z)$ on a 2.5-dimensional Cartesian grid $(x,y,z)$:
\begin{equation}
A_z = \frac{1}{2}\cos{(2x)}+\cos{(y)},
\label{eq:vecpot}
\end{equation}
and the vortex imposes an initial velocity field $\mathbf{v} = (v_x,v_y,0)$ given by
\begin{equation}
v_{\rm x} = -v_{\rm max}\sin{(y)}, \quad v_{\rm y} = v_{\rm max}\sin{(x)},
\label{eq:v}
\end{equation}
where $v_{\rm max} = 0.99c / \sqrt{2}$ is set to ensure that the initial speed $v$ is limited by the speed of light, where the maximum initial Lorentz factor is $\Gamma = (1-v^2/c^2)^{-1/2} = 10$.
The electric field is initialized as $\mathbf{E} = -\mathbf{v} \times \mathbf{B}/c$. 
These settings result in a minimum value of the gas-to-magnetic-pressure ratio $\beta = 2p/b^2 = 10$ and a maximum magnetization $\sigma = b^2 / (\rho h c^2) \approx 0.05$, where $h = 1 + 4p / (\rho c^2) $ is the specific enthalpy for an ideal gas with adiabatic index $\hat{\gamma} = 4/3$ and $b^2$ is the magnetic field strength co-moving with the fluid.
We use a uniform base grid with $128^2$ cells spanning $x,y \in [0, 2\pi]$ with periodic boundary conditions. We then apply subsequently increasing AMR levels to study the convergence of the numerical solution, with each level quadrupling the effective resolution (which is the total number of cells if the highest AMR levels were fully utilised, whereas the actual number of cells is smaller).
The mesh refinement is based on second derivatives in the quantities $B_{z}$, $E_{z}$ and $\rho$ according to the \cite{lohner1987} scheme.
All simulations run till final time $t=10 t_{\rm c}$, with $t_{\rm c}= L/c$ being the light-crossing time, where we assume $L=1$ as the typical length scale of the system.

We consider a set of uniform and constant resistivities, $\eta \in [0, 10^{-5}, 2.5 \times 10^{-5}, 5\times10^{-5}, 10^{-4}, 5\times10^{-4}, 10^{-3}, 5\times10^{-3}]$, chosen such that the corresponding Lundquist numbers $S = L c / \eta \in [\infty, 10^5, 4\times 10^4, 2\times10^4, 10^4, 2\times10^3, 10^3, 2\times10^2]$ range from infinity (ideal GRMHD) to well above and below the threshold for plasmoid formation $S_{th} \approx 10^4$ (\citealt{bhattacharjee2009}; \citealt{uzdensky2010}).

\begin{figure*}
    \centering
     \includegraphics[width=0.32\textwidth]{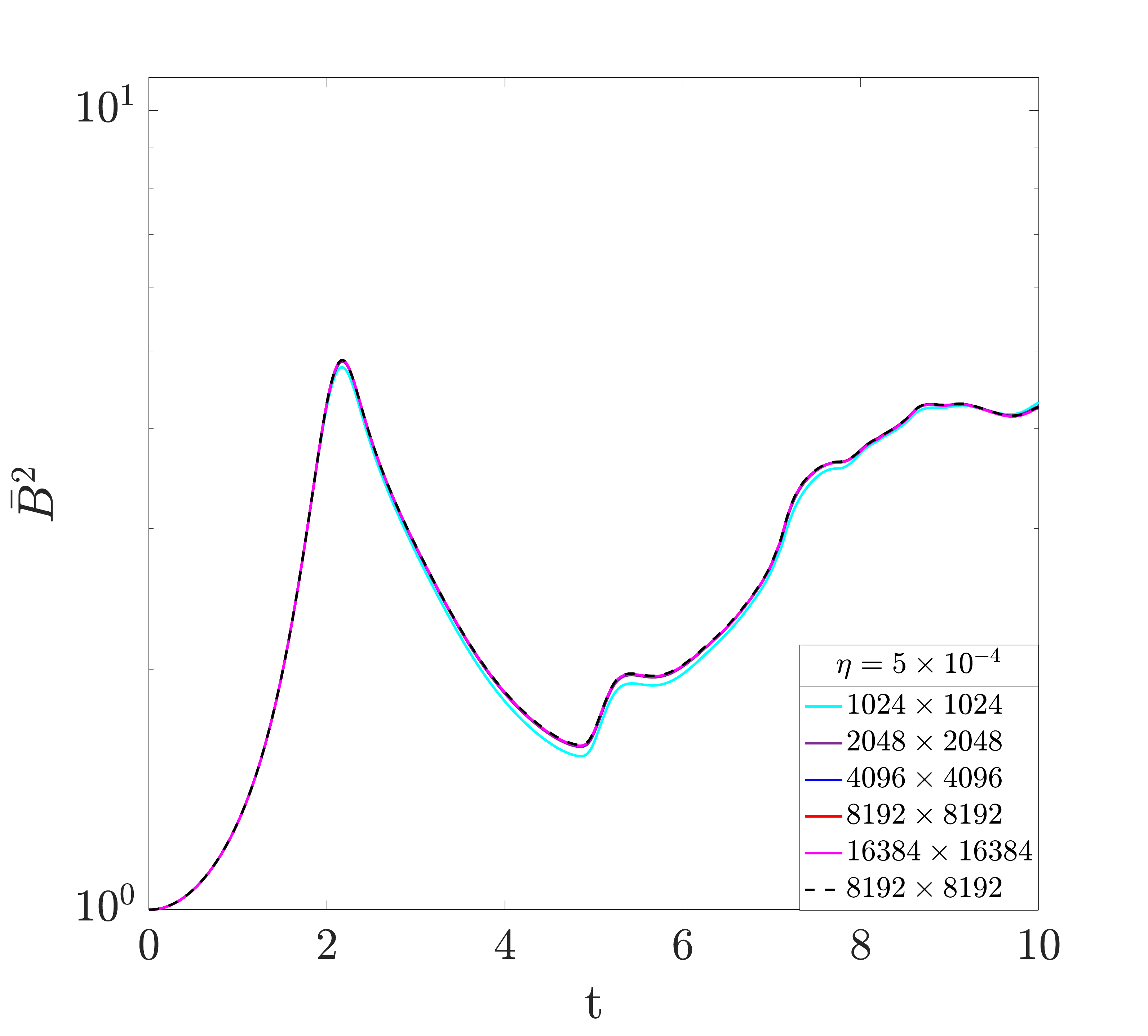}
     \includegraphics[width=0.32\textwidth]{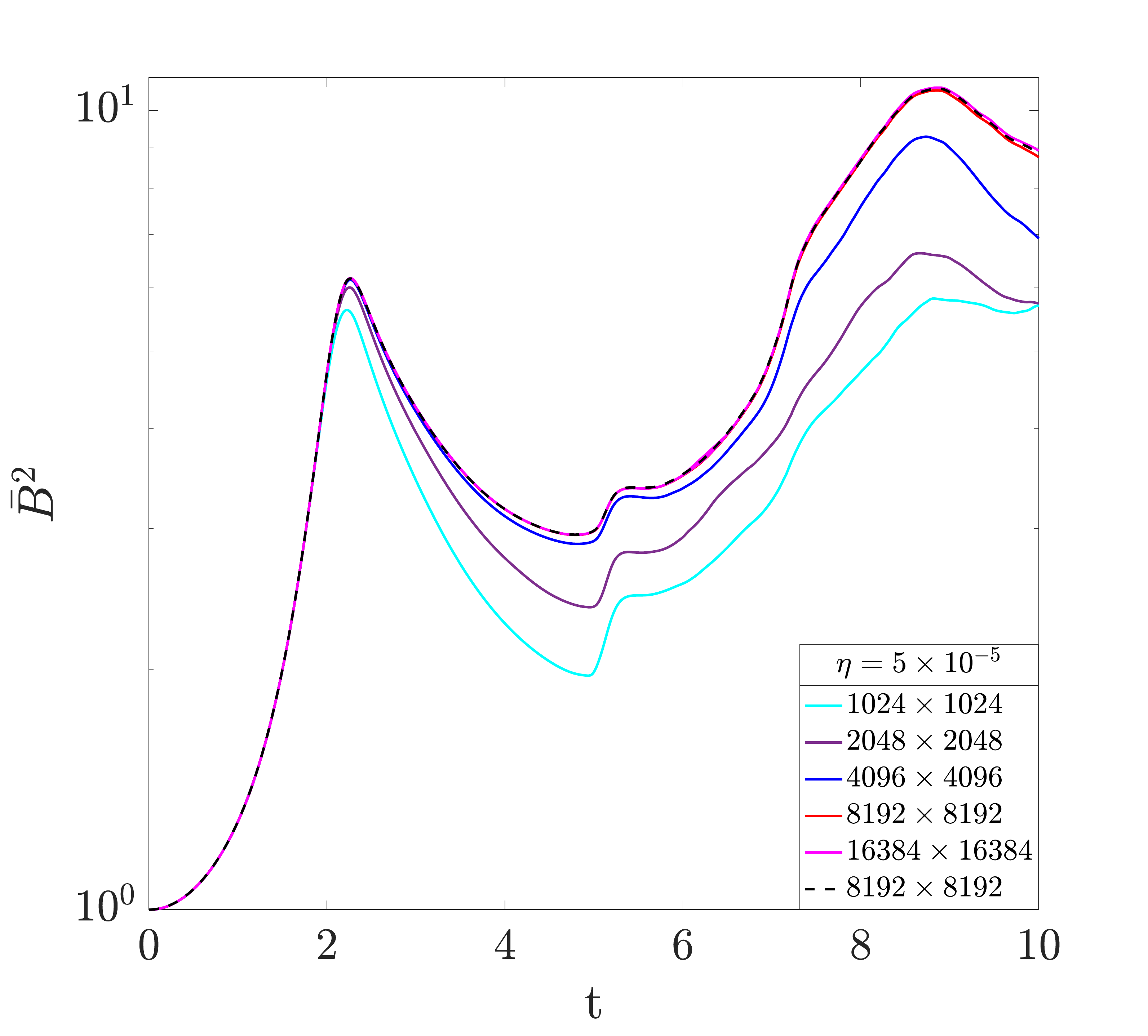}
     \includegraphics[width=0.32\textwidth]{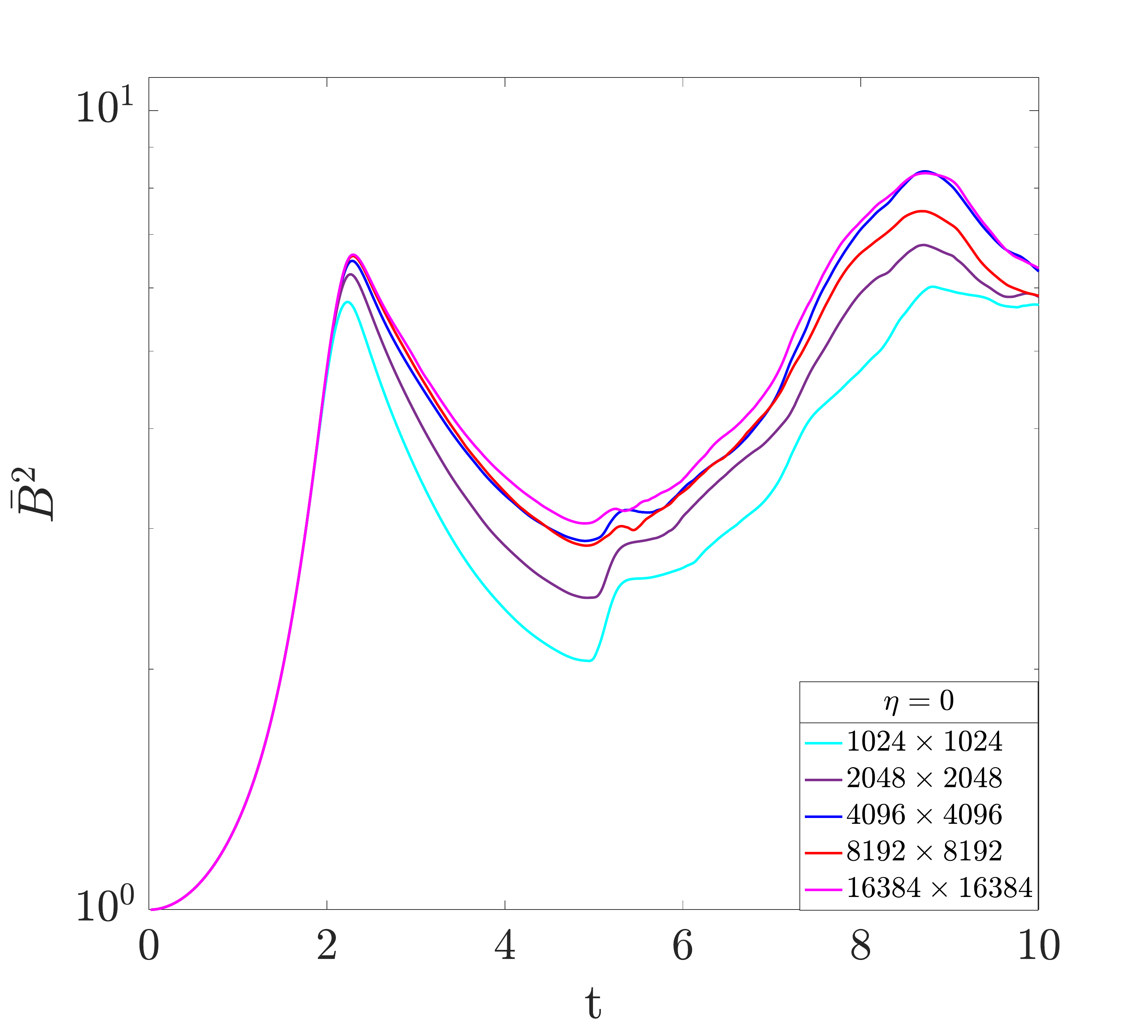}
    \caption{Evolution of the domain-averaged magnetic energy density $\bar{B}^2$ for $S=2\times10^3$ (left), $S=2\times10^4$ (middle), and the ideal case $\eta=0$ (right) for all resolutions considered.}
    \label{fig:OTB2}
\end{figure*}

\subsection{Plasmoid formation}
The initial magnetic field topology of the vortex is characterized by four alternating X-points and magnetic nulls, resulting in two magnetic islands along horizontal lines $y=0$ and $y=\pi$. The vortex motion immediately displaces the left island at $y=\pi$ diagonally upwards and the right island diagonally downwards, resulting in diagonal current sheets. The sheets get compressed, and depending on the Lundquist number, they can become tearing unstable, demonstrating the break-up in a series of plasmoids (see e.g., also \citealt{van_der_Holst_2008}).

For $S\geq 10^4$ the current sheets shrink until they become plasmoid-unstable. In Figure \ref{fig:OTmass} we show the rest-mass density $\rho$ for $S=10^5$ at $t=10 t_{\rm c}$, where plasmoids are recognized as over-dense blobs of plasma in the thin current sheets. 
The current sheet is characterized by an anti-parallel magnetic field configuration resulting in a large out-of-plane component of the current density in the fluid frame,
\begin{equation}
\mathbf{J} = \eta^{-1}\Gamma[\mathbf{E} +\mathbf{v}\times\mathbf{B}/c - (\mathbf{E}\cdot\mathbf{v})\mathbf{v}/c^2],
\label{eq:currentdensity}
\end{equation}
as can be seen in the middle and right panels of Figure \ref{fig:OTplasmoids} for $S=2\times10^4$ and resolution $8192^2$.
For lower Lundquist numbers $S<10^4$ the current sheet does not become thin enough for the plasmoid instability to grow (see e.g., left panel of Figure \ref{fig:OTplasmoids} for $S=2\times10^3$ and resolution $8192^2$).
 
\subsection{Convergence of numerical results}
We claim converged results if the evolution of the domain-averaged magnetic energy density $\bar{B}^2 \equiv \frac{\iint_V B^2 dx dy}{\iint_V dx dy}$ does not change anymore between successive doubling of the number of grid cells per direction over the full domain $V$ (see Figure \ref{fig:OTB2}).
The plasmoid-unstable cases ($S \geq 10^4$) are most demanding and effective resolutions of $8192^2$ cells are needed to resolve the current sheets by at least 10 cells over their width.
For our fiducial case of $S=2\times10^4$ there are no visual differences between runs with resolutions of $8192^2$ and $16384^2$ grid cells (see middle panel of Figure \ref{fig:OTB2}).
For lower resolutions $<8192^2$ the (average) magnetic energy density is underestimated (particularly after $t=2t_{\rm c}$) and the dissipation is governed by numerical resistivity that is larger than the explicit resistivity $\eta = 5\times 10^{-5}$, affecting the energetics, heating, and plasmoid statistics.
We also ensure that the effect of the AMR on the evolution of $B^2$ is negligible compared to a (significantly more expensive) run with uniform resolution of $8192^2$ grid cells (see the dashed black lines in the left and middle panels of Figure \ref{fig:OTB2}).
\begin{figure} 
\centering
\includegraphics[width=0.49\textwidth, clip=true]{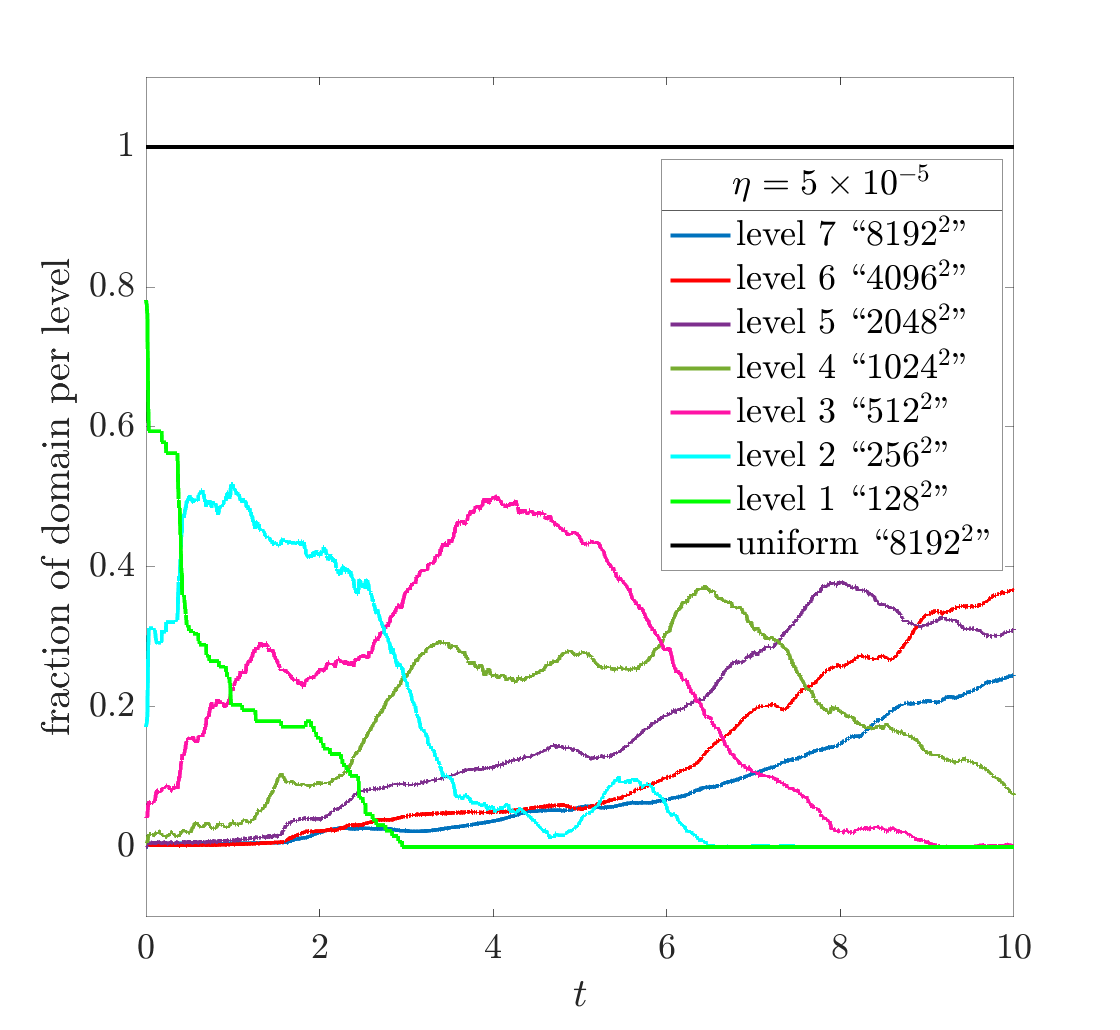}
\caption{The time-evolution of the fraction of the domain covered by the individual AMR levels $n$, calculated as $\mathrm{cells}\times 4^{n_\mathrm{max}-n}$ with $n_\mathrm{max}=7$ levels, corresponding to an effective resolution of $8192^2$ for $S = 2\times10^4$. The black solid line indicates a run with a uniform grid of $8192^2$.}
\label{fig:levels}
\end{figure}
In Fig. \ref{fig:levels} we show the fraction of the domain, calculated as the number of $\mathrm{cells} \times 4^{n_{\rm max}-n}$, with $n$ the AMR level and $n_{\rm max}$ the maximum number of levels, that is covered by the different AMR levels in time for $S = 2\times10^4$. It is clear that the lowest levels, 1 and 2, corresponding to an effective resolution of $128^2$ and $256^2$ respectively, are only important in the first part of the simulation $t < 3 t_{\rm c}$ (i.e., when the plasmoid instability is not yet activated). In the turbulent final stage of the simulation, less than 25\% of the domain is covered by the highest level corresponding to an effective resolution of $8192^2$, and less than 40\% by level 6, corresponding to an effective resolution of $4096^2$, resulting in a major speed-up compared to a uniform resolution of $8192^2$. The highest level is only activated around current sheets, as can also be seen in the right panel of Figure \ref{fig:OTplasmoids}.

We performed an ideal GRMHD (i.e., $\eta=0$) run (see the right panel of Figure \ref{fig:OTB2}), showing that in this case the magnetic energy density evolution, and hence the (numerical) dissipation and heating of the plasma, does not converge even for the highest resolutions considered. Due to the absence of a resistive dissipation scale set by $\eta > 0$ the current sheet shrinks to the grid scale, and is captured by one grid cell over its width for all resolutions considered here.
Notice that the evolution of $B^2$ for $4096^2$ and $16384^2$ grid cells seems comparable for $\eta=0$, whereas the magnetic energy density is clearly lower for the case with $8192^2$ cells.
The lowest resolution ($1024^2$) considered here is generally high enough to capture the essential (non-resistive) effects of the magnetorotational instability (MRI, \citealt{velikhov1959,chandrasekhar1960,balbus1991}), the accretion dynamics, and the Blandford-Znajek jet launching process (\citealt{Blandford1977}) in global disk simulations according to recent studies in ideal GRMHD (\citealt{porth2019,white2019}).
One can see that at resolutions $< 1024^2$, there is hardly a difference between the ideal and $S=2\times10^4$ resistive results (cyan line in middle and right panels of Figure \ref{fig:OTB2}), and the linear growth phase of the magnetic energy density (at $t\lesssim 3 t_{\rm c}$, i.e., where no plasmoids have formed yet) is accurately captured, confirming that numerical resistivity due to the finite grid dominates the magnetic energy dissipation and that higher resolutions are needed to resolve magnetic reconnection and plasmoid formation.
For $S=2\times10^3$ the evolution of the current sheet is easier to capture because the violent tearing instability is not triggered and the explicit resistivity is always larger than the numerical resistivity in this case. The resistive scale is already fully resolved for an effective resolution of $1024^2$ cells (see the left panel of Figure \ref{fig:OTB2}) and the sheet does not shrink below the threshold for plasmoid formation (see the left panel of Figure \ref{fig:OTplasmoids}).

\subsection{Reconnection rate}
\begin{figure}
    \centering
    \includegraphics[width=0.5\textwidth]{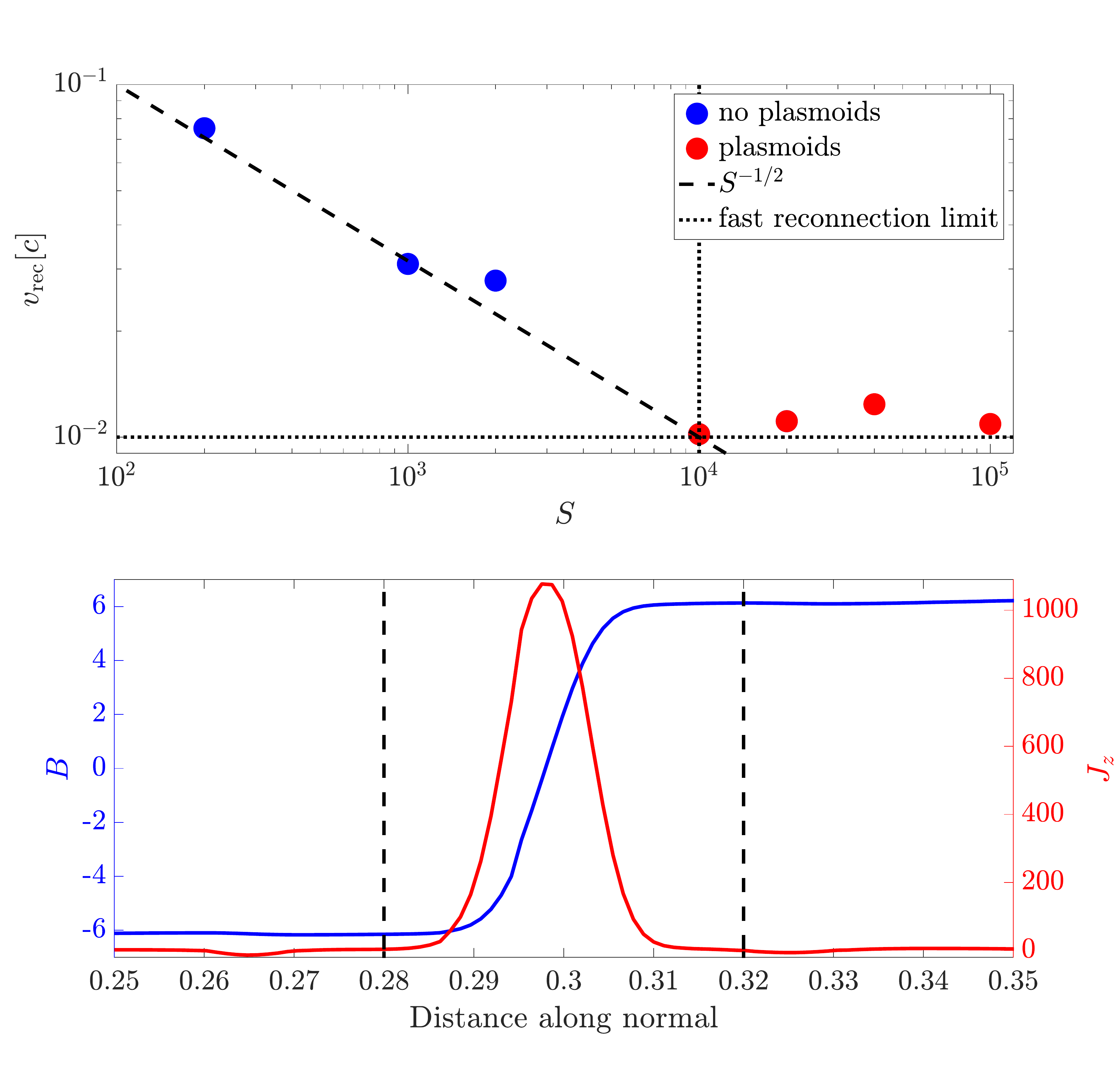}
    \caption{Top panel: Reconnection rate (blue circles) measured as $v_{\rm rec}/c=(v_{\rm up, left}-v_{\rm up, right})/2c$ versus the Lundquist number $S$, showing a Sweet-Parker scaling $v_{\rm rec}/c \sim S^{-1/2}$ (dashed black line) for $S < 10^4$. For $S\geq 10^4$ we observe a ``fast'' reconnection rate (red circles) that is independent of the Lundquist number, as indicated by the dotted back line.
    Bottom panel: In-plane magnetic field $B$ projected along the sheet (blue line) and out-of-plane current density $J_z$ (red line) profiles taken on a slice through the current sheet in the right panel of Figure \ref{fig:OTplasmoids} for $S=2\times10^4$. The magnetic field geometry shows a typical Harris-like current sheet profile. We define the upstream, where we sample the reconnection rate averaged over 10 neighbouring grid cells, to the left and right of the dashed black lines.
    }
    \label{fig:recrateOT}
\end{figure}

We analyze the reconnection rate for all considered Lundquist numbers in the converged runs by measuring the inflow velocity into the current sheet. In order to determine the inflow speed we use the upstream $\mathbf{E}\times\mathbf{B}$-velocity such that $v_{\rm up}/c = E_{z, \rm up} / B_{\rm up}$. 
We take five slices across the current sheet (ensuring that the slice does not cut across a plasmoid) at $t=10 t_{\rm c}$ for all cases with resolutions of $16384^2$ such that even the thinnest sheet for $S=10^5$ is resolved.
We then obtain a profile of $B_{\rm up}$ by projecting the magnetic field along the current sheet and find the location where the profile becomes flat (see the bottom panel of Figure \ref{fig:recrateOT} for $S=2\times10^4$).
We measure both the upstream electric and magnetic fields by averaging over 5 locations in the upstream on the slices on both sides of the sheet.
We account for the bulk flow velocity ($v_{\rm bulk}\ll c$ such that $\Gamma_{\rm bulk} = 1$) of the vortex by defining the speed on the left of the sheet as $v_{\rm up, left}/c = (v_{\rm bulk} + v_{\rm in}$)/c and on the right of the sheet as $v_{\rm up, right}/c = (v_{\rm bulk} - v_{\rm in})/c$, such that  $v_{\rm rec}/c=(v_{\rm up, left}-v_{\rm up, right})/2c$. Assuming that the outflow $v_{\rm out} = v_{A} \approx c$, this directly yields the reconnection rate $v_{\rm rec}/c = v_{\rm in} / v_{\rm out} = v_{\rm in}/c$.
We confirmed that locally, in the upstream region of the current sheet, $\sigma = b^2 / (\rho h c^2) \approx 8$ such that the Alfv\'{e}n speed around the sheet is $v_A = c(\sigma/(\sigma+1))^{1/2} \approx c$ and that the reconnection is relativistic.
In the top panel of Figure \ref{fig:recrateOT} we observe a Sweet-Parker scaling $v_{\rm rec}/c \sim S^{-1/2}$ for $S <10^{-4}$ (indicated by the blue circles and the dashed black line) and plasmoid-dominated ``fast'' reconnection with a rate independent of the Lundquist number of $v_{\rm rec}/c \approx 0.01$ for $S \geq 10^4$ (indicated by the red circles and the dotted black line).

\section{Magnetic reconnection and plasmoid formation in black-hole accretion flows}
\begin{figure}
    \centering
    \includegraphics[width=0.5\textwidth]{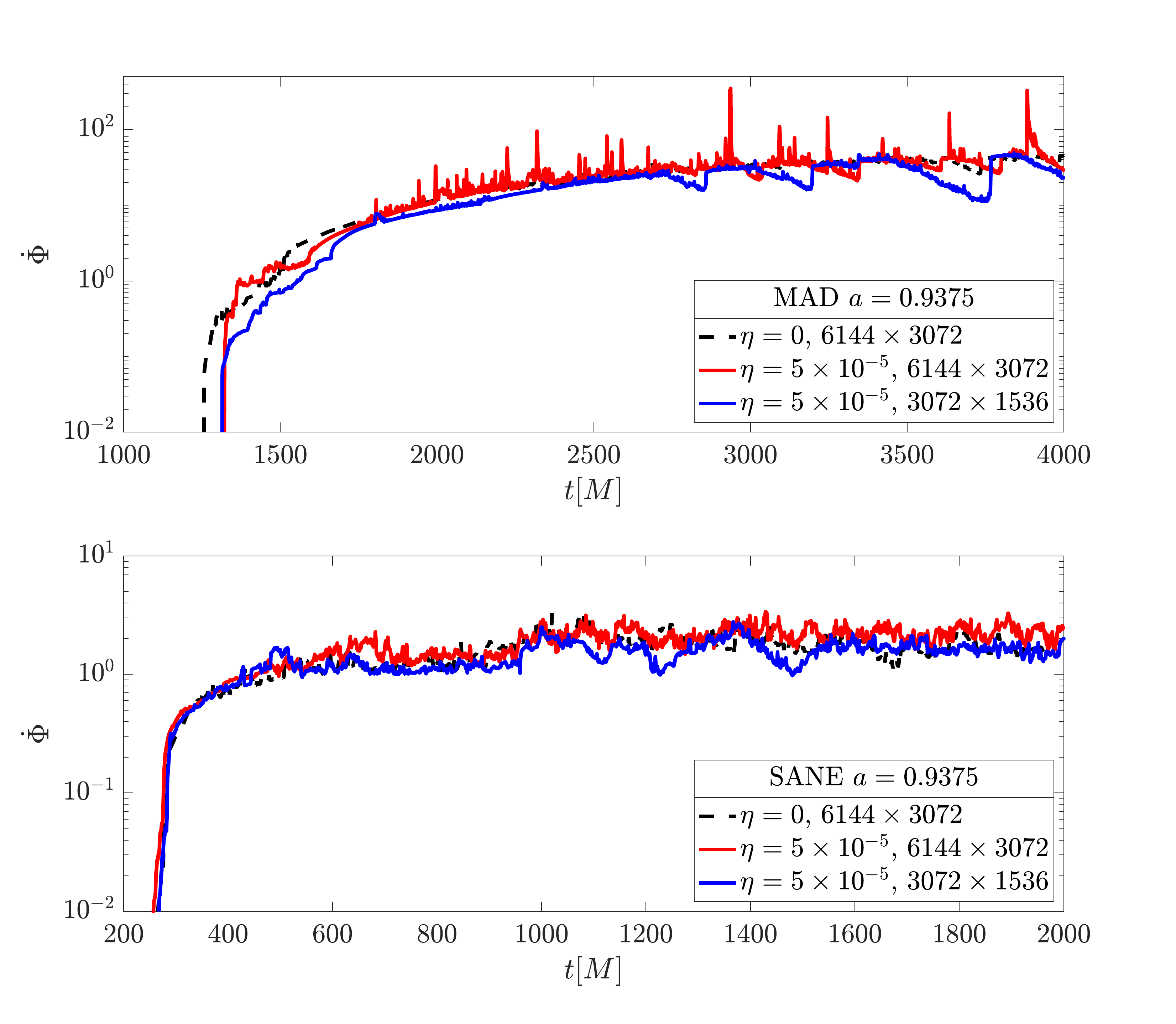}
    \caption{Time-evolution of the magnetic flux through the horizon $\dot{\Phi}$ for a MAD run (top) and a SANE run (bottom) for $\eta=0$ (dashed black lines) and $\eta=5\times10^{-5}$ (solid red lines). The ideal and resistive results at a resolution of $6144\times3072$ are in agreement and the global accretion dynamics are unaffected by resistivity. We also show the resistive run at a resolution that is twice smaller per direction, $3072\times1536$ (blue lines), to show convergence of the global accretion dynamics.}
    \label{fig:mdot}
\end{figure}
It is much harder to localize and track the formation of current sheets in realistic black-hole accretion flows in a larger domain and for a longer period because of the effects of the more complicated global dynamics governed by the central object, and due to the turbulence induced by the MRI.
Both the evolution of accretion flows and the formation of current sheets therein strongly depend on the magnetic field geometry. 
We model an accretion disk around a rotating black hole varying the initial conditions to study current sheet formation in different scenarios of magnetic field geometry.
In the Magnetically Arrested Disk (MAD, \citealt{Igumenshchev2003,narayan2003}) scenario the MRI and subsequent turbulence in the inner accretion disk are suppressed due to large-scale magnetic flux (see e.g., \citealt{white2019}).
In axisymmetric simulations as considered here, the arrested inflow is regularly broken by frequent bursts of accretion, allowing for a macroscopic equatorial current sheet to form and break in a periodic fashion. In a full 3D setup magnetically buoyant structures are interchanged with less-magnetized dense fluid (\citealt{Igumenshchev2008,white2019}), resulting in a magnetic Rayleigh-Taylor instability (\citealt{kruskal1954}) potentially sourcing interchange-type magnetic reconnection.
In the Standard And Normal Evolution (SANE, \citealt{Narayan2012,sadowski2013}) state a fully turbulent accretion disk can develop due to a smaller magnetic flux (see e.g., \citealt{porth2019}), and current sheets can ubiquitously form and interact with the turbulent flow. 
Polarized synchrotron radiation observed by the Event Horizon Telescope (EHT, \citealt{EHTpaper1}) can probe the field line structure at event-horizon scales and put tighter constraints on the magnetization and address whether the accretion is in a SANE or a MAD state (\citealt{EHTpaper5}). 
\begin{figure*}
    \centering
    \includegraphics[width=0.275\textwidth,trim= 13cm 1cm 14.7cm 2cm, clip=true]{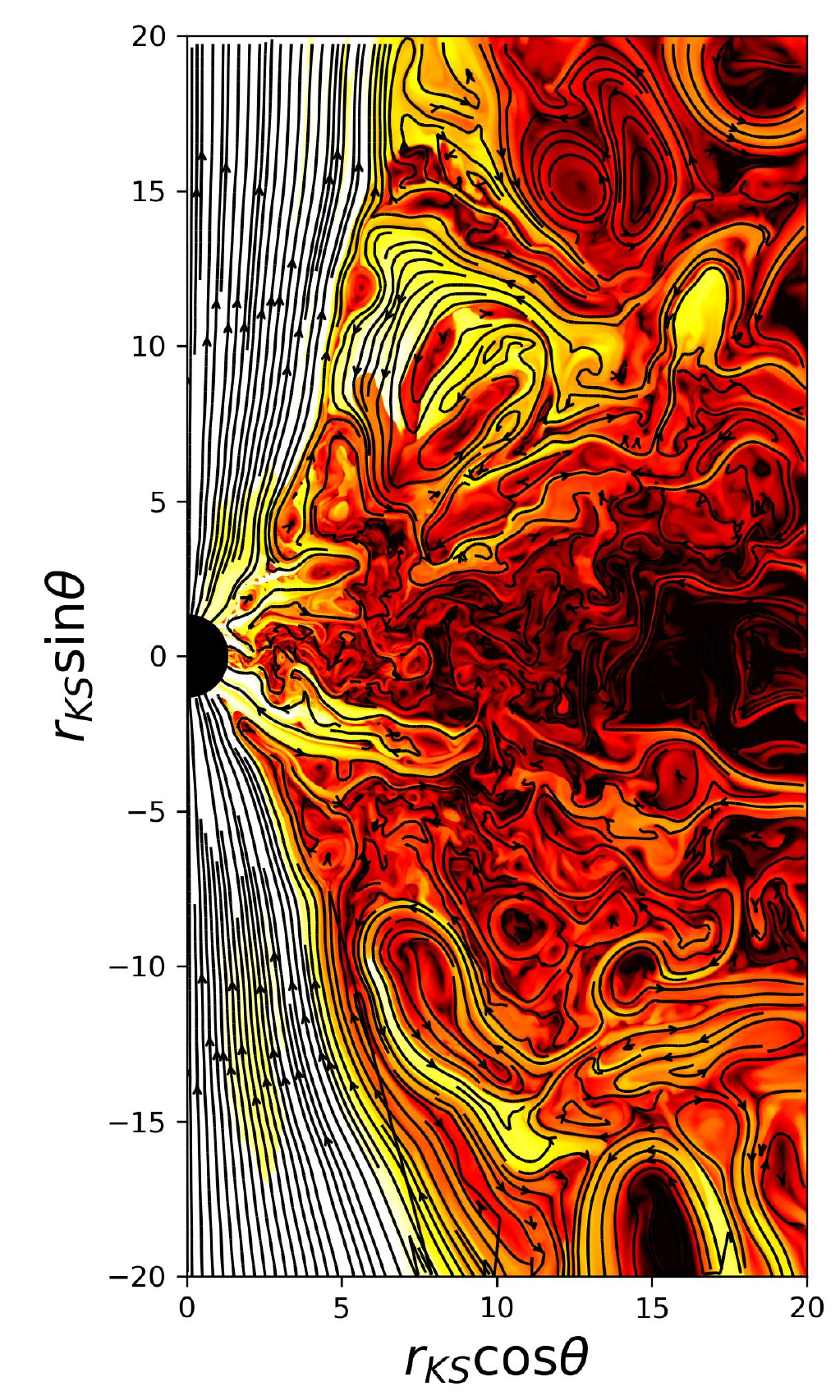}
    \includegraphics[width=0.2167\textwidth,trim= 15.2cm 1cm 14.7cm 2cm, clip=true]{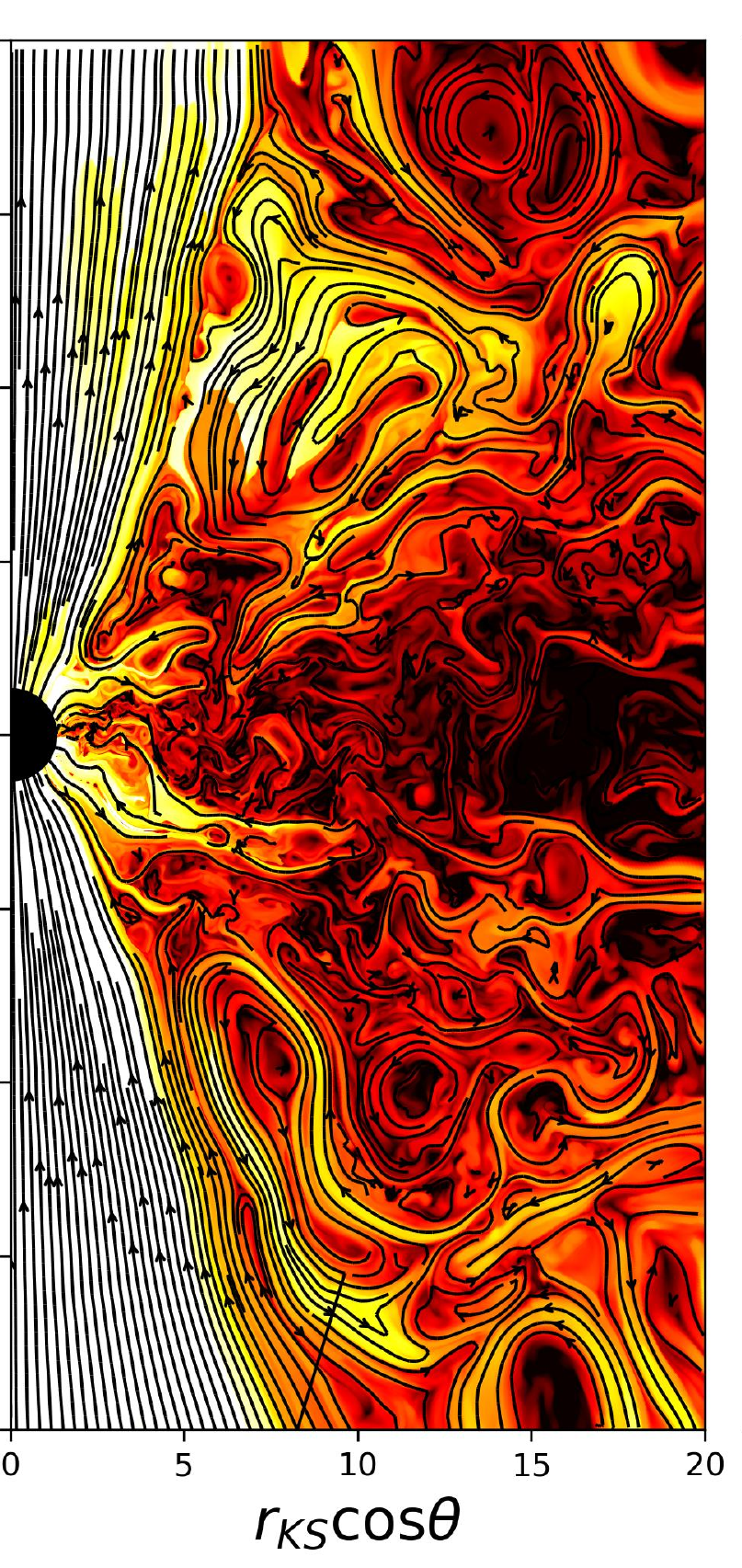}
    \includegraphics[width=0.2167\textwidth,trim= 15.2cm 1cm 14.7cm 2cm, clip=true]{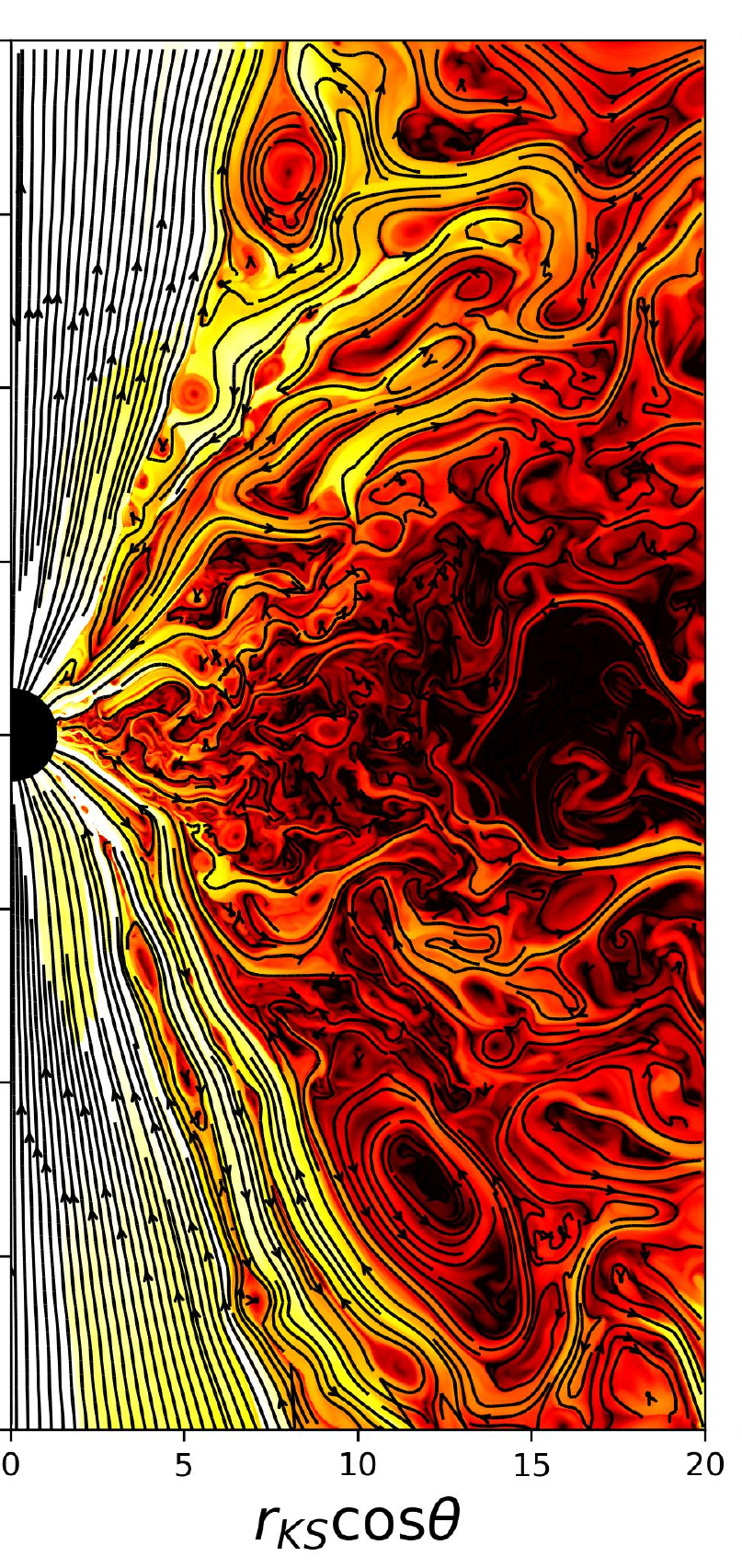}
    \includegraphics[width=0.275\textwidth,trim= 15.2cm 1cm 12.5cm 2cm, clip=true]{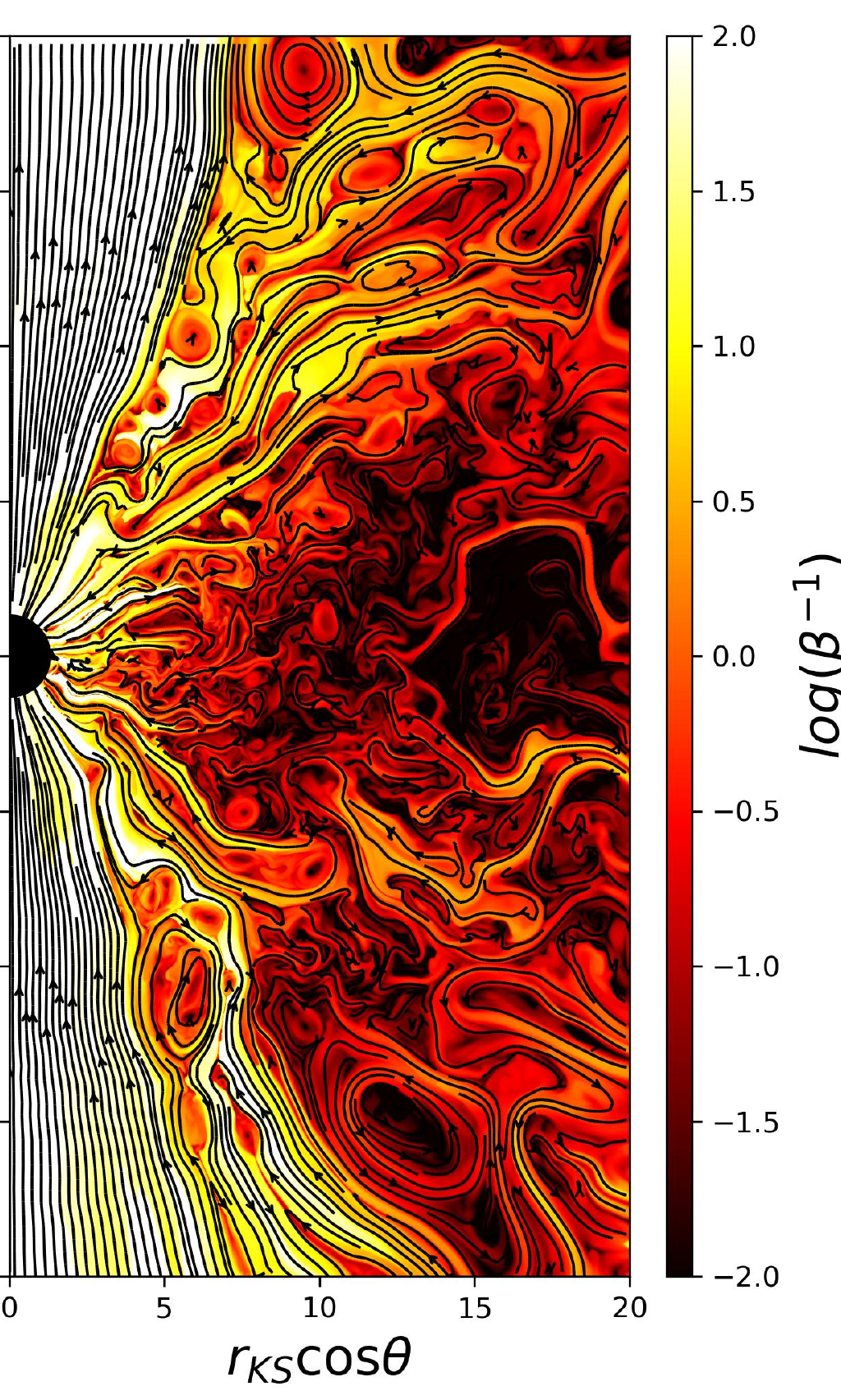}
    \caption{$\beta^{-1}=b^2/(2p)$ at four typical times $t = [1460, 1480,1540, 1580] r_g/c$ (from left to right) during quasi-state-state phase of accretion in the SANE configuration. Magnetic field lines plotted are on top as solid black lines. In the bottom half one can detect the accretion of a magnetic flux tube at $r_{KS}\cos\theta = x \approx 6r_g, r_{KS}\sin\theta = y \approx -8r_g$ (left two panels) that inflates, opens up and becomes tearing unstable (third panel) after it connects to the black hole, and produces copious plasmoids coalescing into large-scale structures at $x\approx5r_g,y\approx-10r_g$ (fourth panel) with a typical size of about one Schwarzschild radius. In the top half of the panels a similar process can be seen at $x\approx7r_g,y\approx13r_g$ in the second and third panel, also resulting in a large-scale plasmoid at $x\approx9r_g,y\approx18r_g$ in the fourth panel.}
    \label{fig:SANE154betaevolution}
\end{figure*}

Here, we consider both the SANE and MAD scenarios, studying whether forming current sheets can become tearing-unstable and produce macroscopic plasmoids before breaking up. 
We also investigate how large these plasmoids can grow and whether they are bounded to the disk or expelled from the disk, e.g., along the jet's sheath.

\subsection{Numerical setup}
We consider both MAD and SANE magnetic field configurations around a Kerr black hole, for three representative values of the dimensionless spin parameter $a \in [-0.9375;0;0.9375]$, corresponding to near-extremal retrograde Kerr, Schwarzschild, and near-extremal prograde Kerr black holes, respectively.
We use geometrized units with gravitational constant, black-hole mass, and speed of light $G = M = c = 1$; such that length scales are normalized to the gravitational radius $r_g = GM/c^2 $ and times are given in units of $r_g/c$. 
We employ spherical Kerr-Schild coordinates, where $r$ is the radial coordinate, $\theta$ and $\phi$ are the poloidal and toroidal angular coordinates, respectively, and $t$ is the temporal coordinate.
We start our simulations from a torus in hydrodynamic equilibrium (\citealt{fishbone1976}) threaded by a single weak poloidal magnetic field loop, defined by the vector potential
\begin{equation}
    A_{\phi} \propto \max(q, 0), 
\end{equation}
where the function $q$ is set to obtain a large torus resulting in a MAD state:
\begin{equation}
    q = \frac{\rho}{\rho_{\rm max}}\left(\frac{r}{r_{\rm in}}\right)^3\sin^3\theta\exp\left(-\frac{r}{400}\right)-0.2,
\end{equation}
with an inner radius $r_{\rm in} = 20 r_g$ and the density maximum $\rho_{\rm max}$ is located at $r_{\rm max} = 41 r_g$ for $a=0.9375$, $r_{\rm max} = 42 r_g$ for $a=-0.9375$, and $r_{\rm max} = 41.4 r_g$ for $a=0$ to start from an initial torus of similar size.
For a SANE state we set a smaller torus with
\begin{equation}
    q = \frac{\rho}{\rho_{\rm max}}-0.2,
\end{equation}
where the inner radius $r_{\rm in} = 6 r_g$ and the density maximum $\rho_{\rm max}$ is located at $r_{\rm max} = 12 r_g$ for $a=0.9375$, $r_{\rm max} = 22 r_g$ for $a=-0.9375$, and $r_{\rm max} = 15 r_g$ for $a=0$ to start from an initial torus of similar size.
In both cases the magnetic field strength is set such that $\beta = 2 p_{\rm max}/b^2_{\rm max} = 100$. Plasma-$\beta$ and the magnetization $\sigma=b^2/(\rho c^2)$ for a cold (i.e., $p\ll\rho$) plasma are defined using the magnetic field strength $b^2$ co-moving with the fluid (see \citealt{ripperda2019b} for a definition).
We set an atmospheric rest-mass density and pressure as $\rho_{\rm atm} = \rho_{\rm min}r^{-3/2}$ and $p_{\rm atm} = p_{\rm min}r^{-5/2}$ where $\rho_{\rm min} = 10^{-4}$ and $p_{\rm min} = 10^{-6}/3$. 
We apply floors on rest-mass density, pressure, and Lorentz factor $\Gamma < \Gamma_{\rm max}=20$ such that the magnetization $\sigma = b^2/(\rho c^2) < \sigma_{\rm max}=100$ and $\beta^{-1} = b^2 / 2p < \beta^{-1}_{\rm max} = (10\sigma_{\rm max})^{(\hat{\gamma}-1)}$.
We adopt an equation of state for a relativistic ideal gas with an adiabatic index of $\hat{\gamma} = 4/3$. 
The equilibrium fluid pressure is perturbed to trigger the MRI as $p = p_{\rm eq}(1+X_p)$ with a random variable uniformly distributed between $X_p \in [-0.02;0.02]$.

We model dissipation in the accretion flow by assuming a small, uniform, and constant resistivity $\eta = 5\times10^{-5}$ in the set of GRRMHD equations (\citealt{ripperda2019b}), resulting in a large Lundquist number $S = L c / \eta \sim 2\times 10^5$ that is well above the plasmoid threshold, where we assume the typical length scale to be the length of a current sheet $L = L_{\rm{sheet}} \sim \mathcal{O}(10 r_g)$, and the speed of light $c$ as the typical velocity.
A small resistivity allows us to capture both the global near-ideal accretion dynamics (\citealt{ripperda2019b}) and the fast reconnection resulting in plasmoid formation in localized thin current sheets (\citealt{ripperda2019}). 

\subsection{Convergence of numerical results}
We show in Figure \ref{fig:mdot} that both MAD and SANE configurations in ideal and resistive runs reach a quasi-steady-state after approximately $2500 r_g/c$ and $500 r_g/c$ respectively, as indicated by the magnetic flux through the horizon:
\begin{equation}
 \dot{\Phi}:=\frac{1}{2}\int_0^{2\pi}\int^\pi_0|B^r|\sqrt{-\gamma}d\theta d\phi,
\end{equation}
where $\gamma$ is the determinant of the spatial part of the metric.

Figure \ref{fig:mdot} shows the magnetic flux for the highest resolution considered, $6144\times3072$ cells, compared to a resolution of $3072\times1536$, confirming that quasi-steady-state phase of accretion is independent of the resolution in our simulations. We confirm convergence of the thinning process of the current sheets by restarting from the quasi-steady-state increasing the resolution up to $12288\times6144$, and measure the cells per thickness of the sheets. In all considered cases the sheets are captured by 8 or more cells over their widths, which is discussed in section \ref{sec:SANE}.
\begin{figure*}
    \centering
    \includegraphics[width=0.43\textwidth,trim= 13cm 1cm 12.5cm 2cm, clip=true]{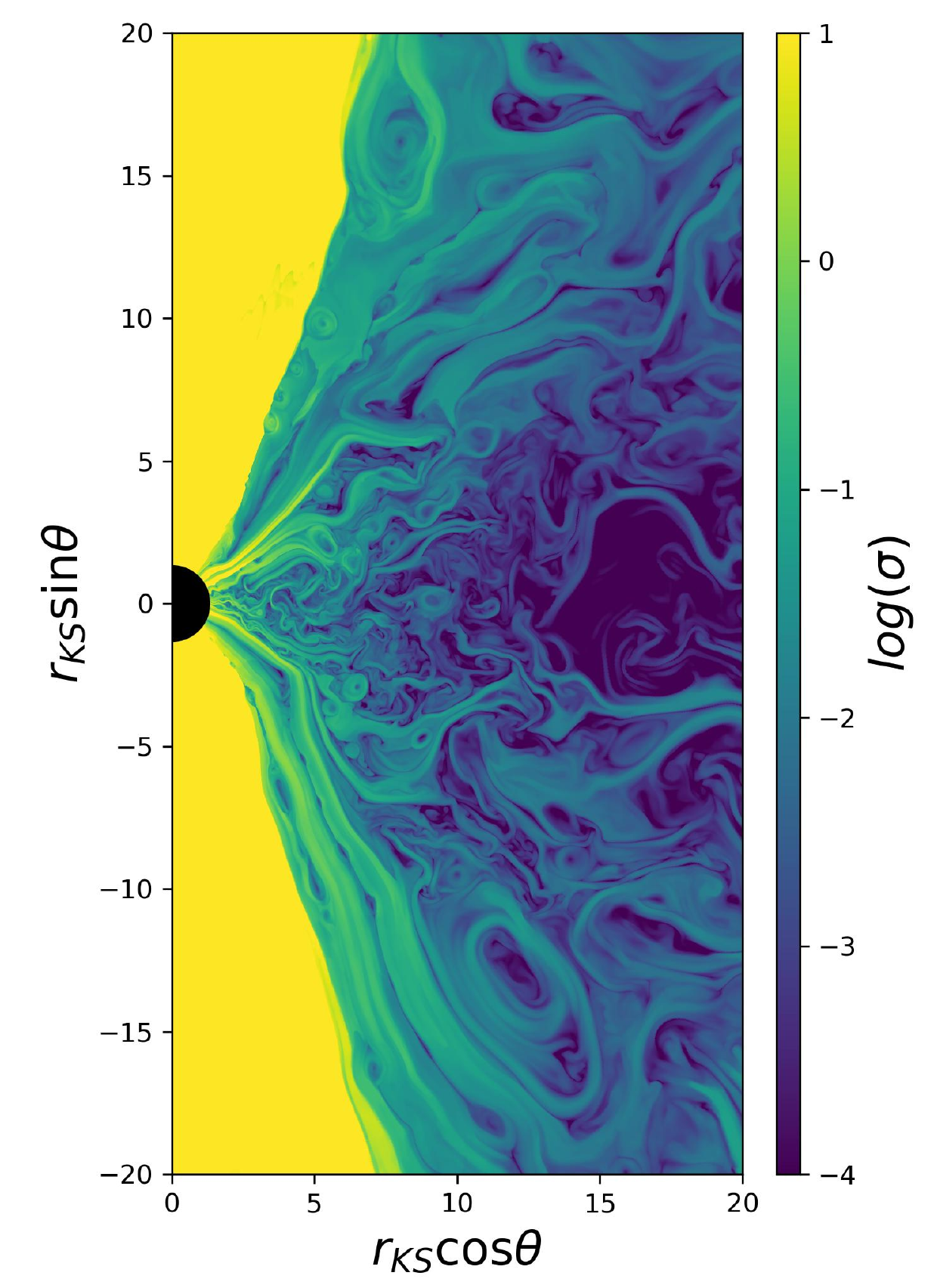}
    \includegraphics[width=0.43\textwidth,trim=13cm 1cm 12.5cm 2cm, clip=true]{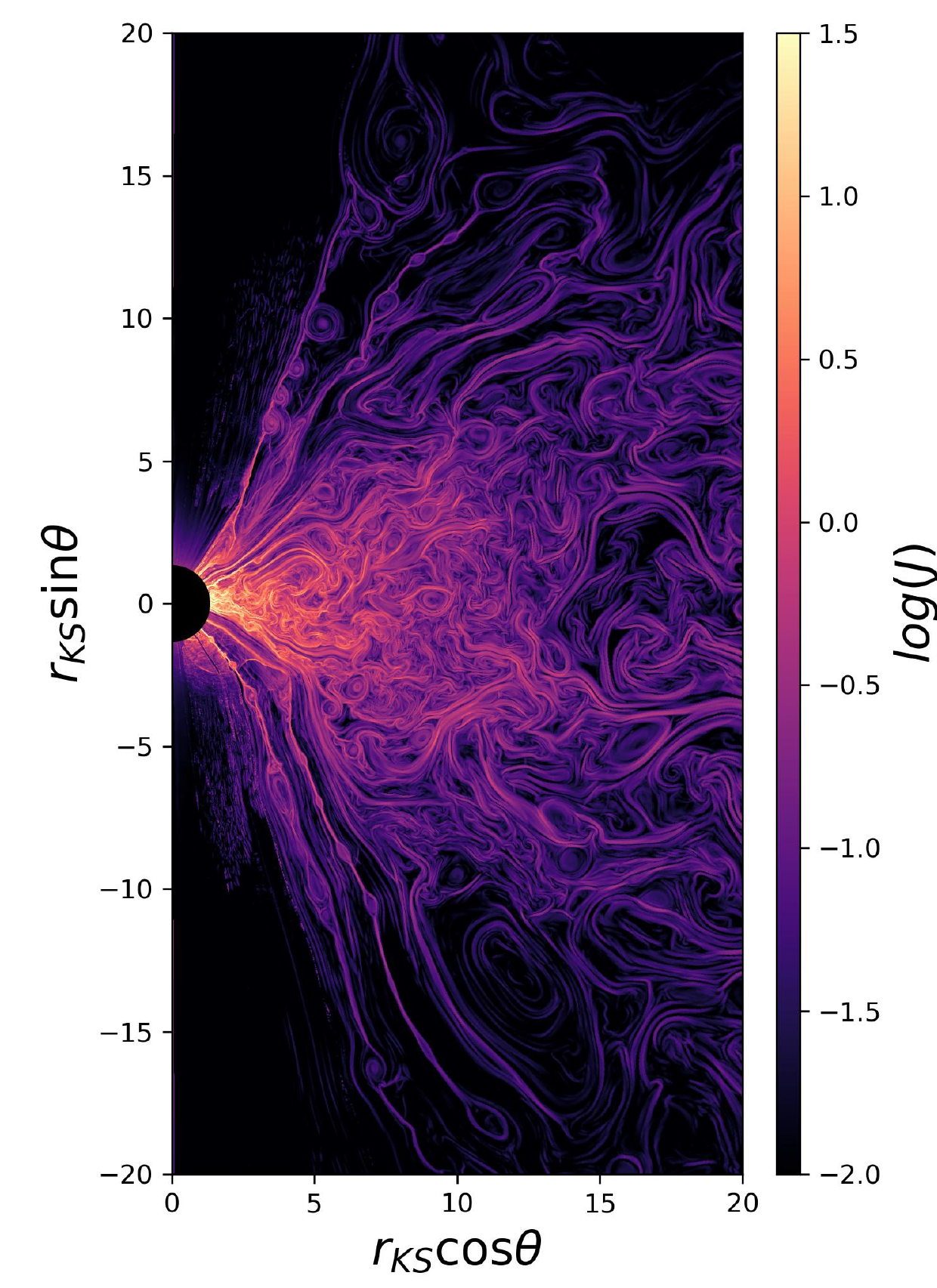}
    
    \includegraphics[width=0.43\textwidth,trim= 13cm 1cm 12.5cm 2cm, clip=true]{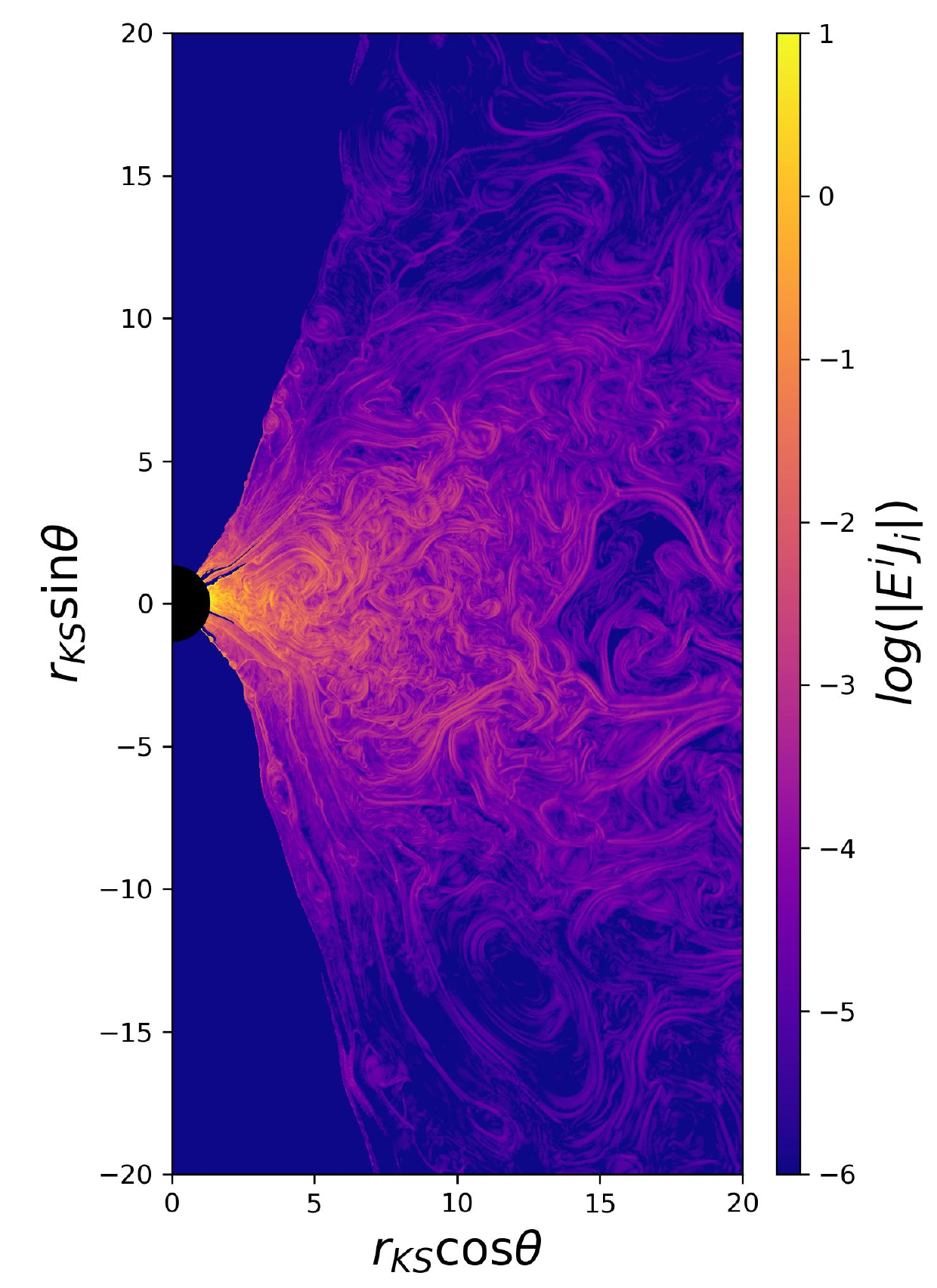}
    \includegraphics[width=0.43\textwidth,trim= 13cm 1cm 12.5cm 2cm, clip=true]{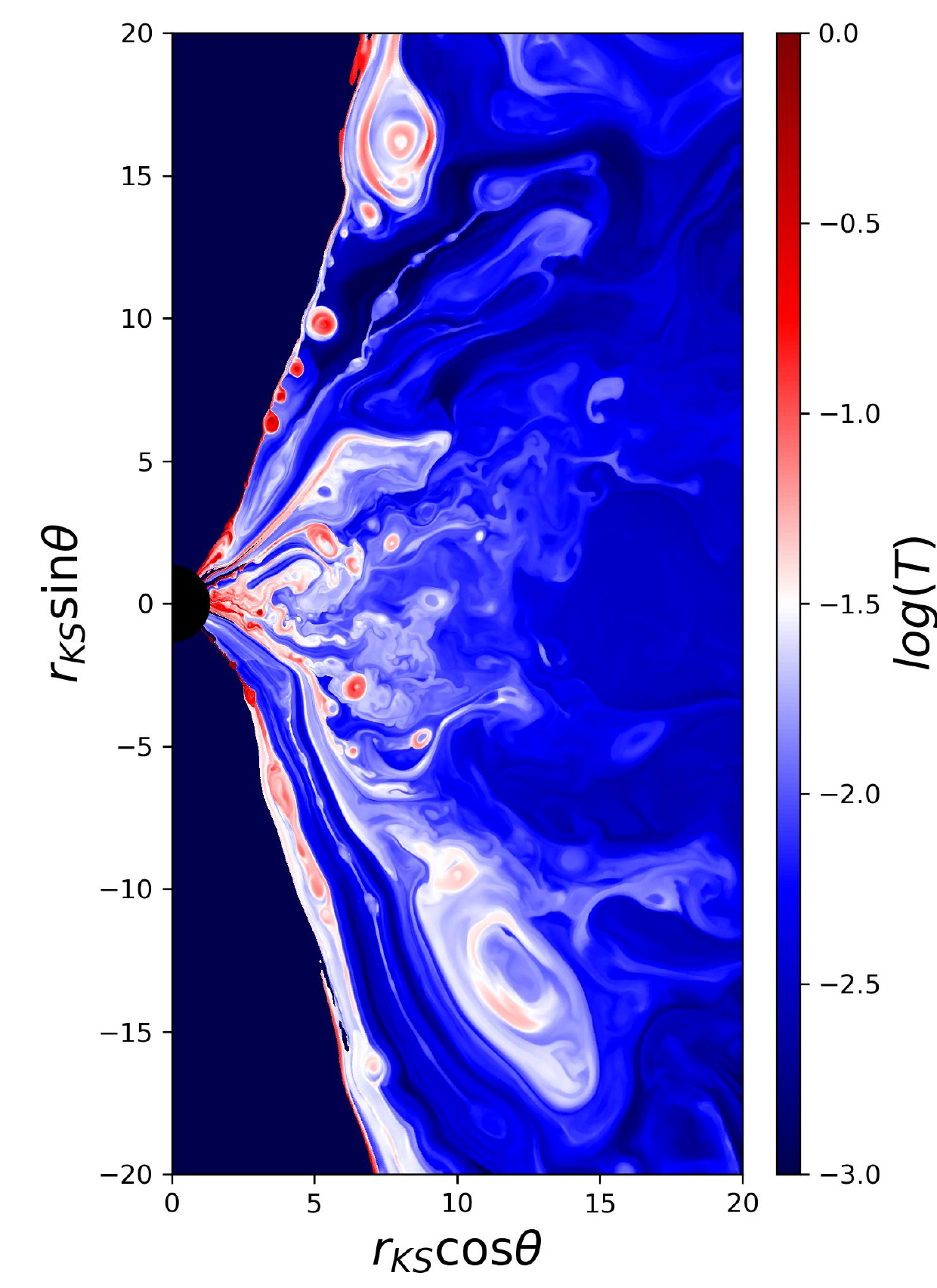}
    \caption{Representative SANE snapshot in quasi-steady-state phase of accretion at $t = 1540 r_g/c$. top-left: Magnetization $\sigma$ showing that current sheets along the jet's sheath are in the relativistic regime $\sigma \gg 1$, whereas in the disk they are in the transrelativistic regime $\sigma \lesssim 1$.
    top-right: The thin tearing-unstable reconnection layers are indicated by a strong current density.
    bottom-left: Plasmoids in the current sheets are heated by Ohmic heating close to the event horizon.
    bottom-right: Plasmoids both in the disk and along the jet's sheath are heated up to relativistic temperatures $T =p/\rho \sim 1$.}
    \label{fig:SANE154}
\end{figure*}

\subsection{Plasmoid formation in the SANE model}
\label{sec:SANE}
Current sheets are expected to form above the disk or in the jet's sheath where magnetic flux tubes are twisted by global shearing motion. These structures can inflate while they accrete onto the black hole and get thinner after which their magnetic energy is dissipated close to the event horizon through reconnection. The magnetic energy is released into heat and bulk motion of the plasmoids that can either fall into the black hole or get ejected. This process has been studied in the force-free paradigm, assuming the plasma to be infinitely magnetized (\citealt{Parfrey_2014,Yuan_2019b,Yuan_2019a,mahlmann2020}). 
Additionally, when the net magnetic flux in the accretion disk is relatively small, turbulence resulting from the MRI can produce magnetic fields with alternating polarities prone to reconnection (\citealt{davis2010,zhu2018}). MHD turbulence is known to intermittently form large plasmoid-unstable current sheets (\citealt{Zhdankin_2013,Zhdankin_2017,dong2018}) and the plasmoid instability can significantly modify the turbulent MHD cascade at relatively small scales and high Lundquist numbers $S\sim 10^6$ (\citealt{boldyrev2017,Comisso_2018,dong2018}).
\begin{figure}
    \centering
    \includegraphics[width=0.35\textwidth,trim= 13.8cm 1cm 12.5cm 2cm, clip=true]{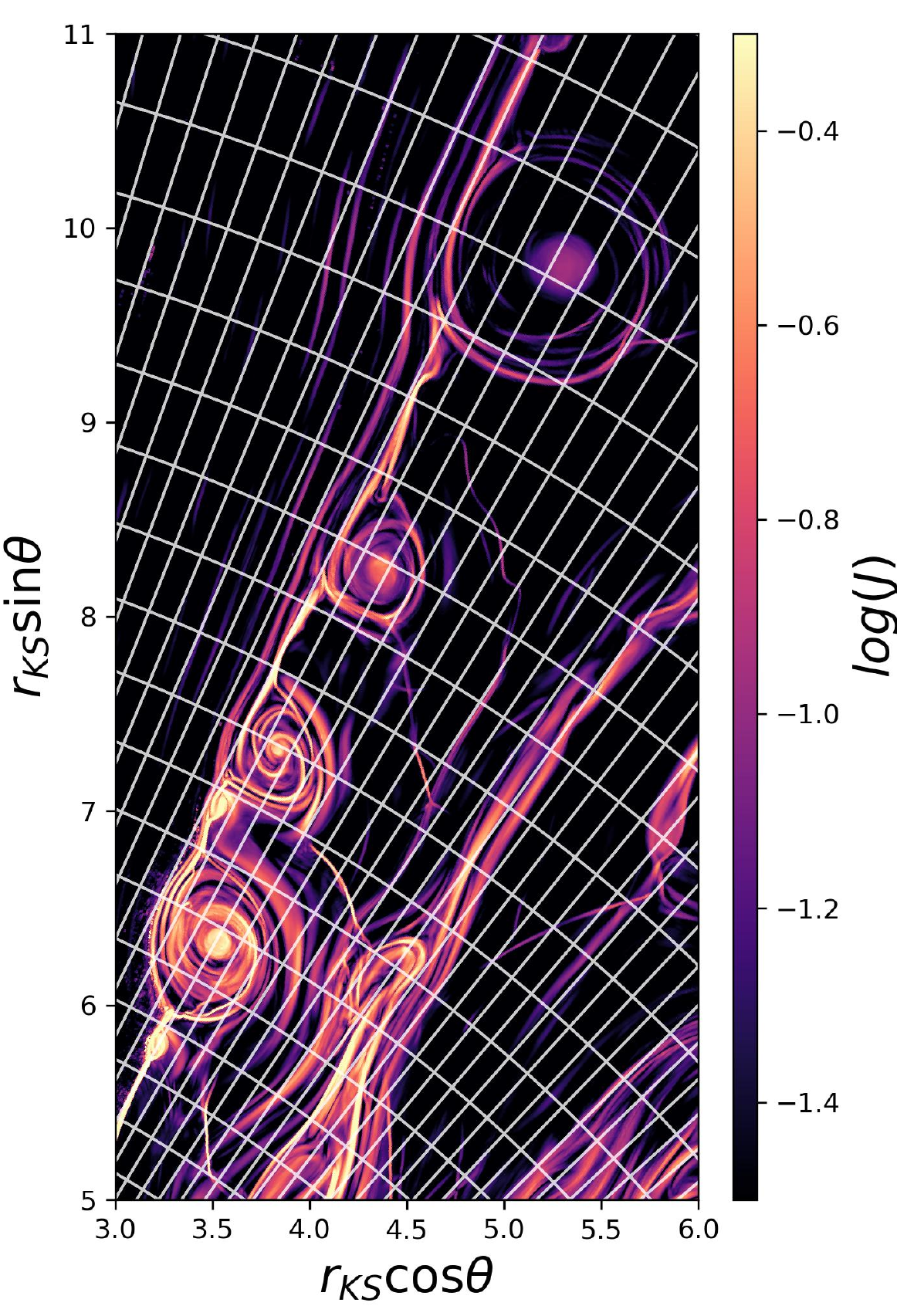}
    \caption{Zoom into a current sheet (see top-right panel of Figure \ref{fig:SANE154}), showing plasmoids that are advected along the jet's sheath and reached a typical size of the order of a few Schwarzschild radii through coalescence. The over-plotted grid-block structure (white rectangles) shows that each current sheet is captured by approximately 10 cells over its width. Each block represents $64 \times 32$ cells.}
    \label{fig:SANE162}
\end{figure}

In Figure \ref{fig:SANE154betaevolution} we observe both processes in a SANE configuration and detect current sheets in the disk and along the jet's sheath, indicated by a small $\beta^{-1}$ and anti-parallel field lines. In the left-most two panels at $t = 1460 r_g/c$ and $t = 1480 r_g/c$ a large flux tube falls onto the black hole in the left bottom half at approximately $x=r_{\rm KS}\cos\theta \approx 8 r_g$, $y=r_{\rm KS}\sin\theta \approx -8 r_g$. 
In the third panel, at $t = 1540 r_g/c$, the flux tube has both its footpoints attached to the black hole, it opened up after it inflated and became thin enough to be tearing unstable and form multiple plasmoids (\citealt{Parfrey_2014} observe a similar process). In the fourth panel, at $t = 1580 r_g/c$, the plasmoids that are advected away from the black hole along the jet's sheath have formed a large structure at $x \approx 6r_g, y\approx-10r_g$ through coalescence. 
At $x\approx 9r_g, y\approx17r_g$ a similar process occurs in the third panel and a large plasmoid has formed through mergers of multiple smaller plasmoids that can be seen in the left panels. In the first and second panel one can detect a flux tube with one footpoint connected to the black hole, that is then twisted by the shear flow, becomes thinner and inflates, at $x\approx3r_g,y\approx2r_g$, after which it ejects plasmoids into the accretion disk in the fourth panel at $x\approx6r_g,y\approx2r_g$. A similar process occurs in the first panel for a loop with one footpoint on the black hole at $x\approx 7r_g,y\approx-2r_g$. The large-scale plasmoids form on a time scale of $\sim \mathcal{O} (100 r_g/c)$ growing to circular objects with a radius $\gtrsim \mathcal{O}(r_g)$, indicating a reconnection rate of $\approx 0.01c$. These structures eventually break up and lose coherence due to interaction with the ambient flow.
\begin{figure*}
    \centering
    \includegraphics[width=0.275\textwidth,trim= 13cm 1cm 14.7cm 2cm, clip=true]{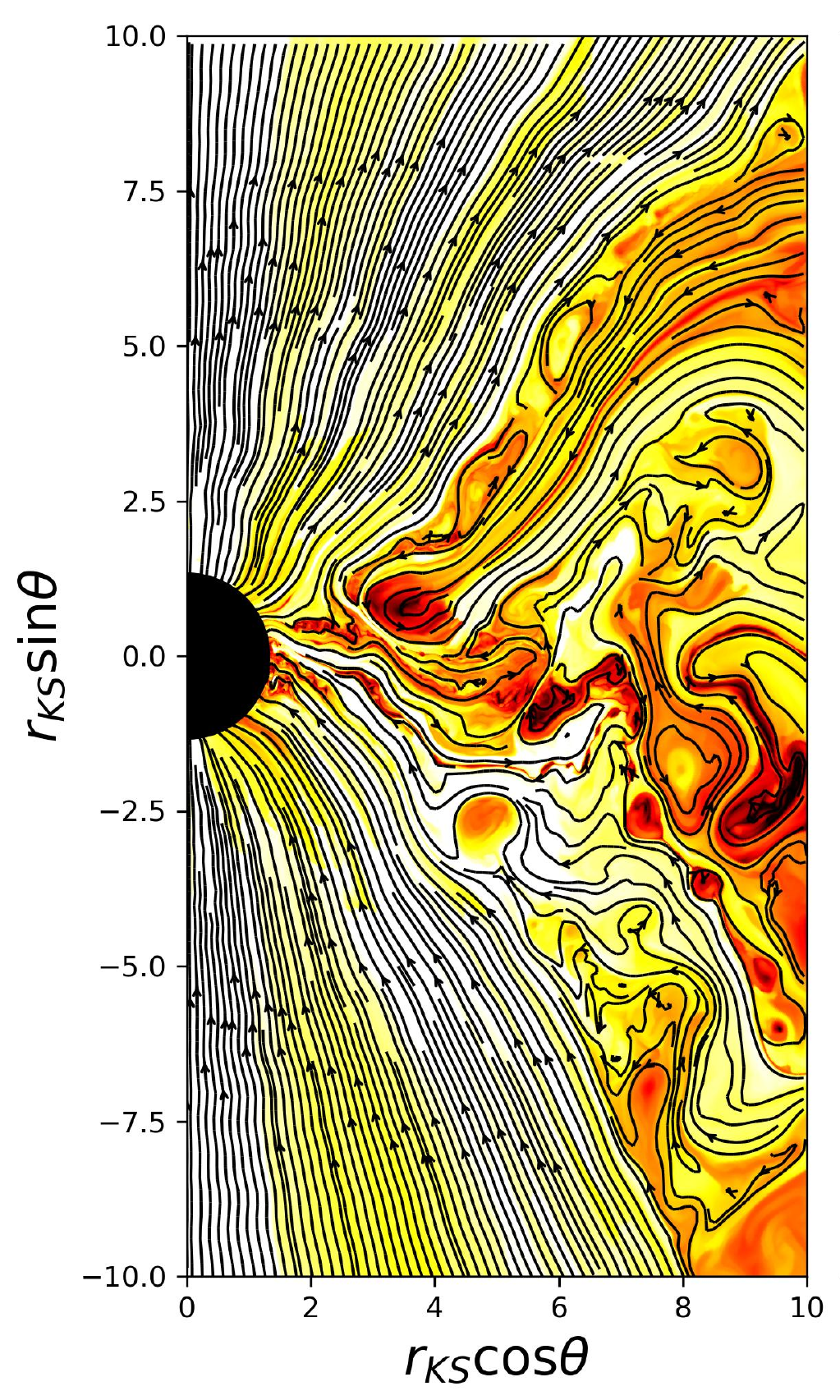}
    \includegraphics[width=0.2167\textwidth,trim= 15.2cm 1cm 14.7cm 2cm, clip=true]{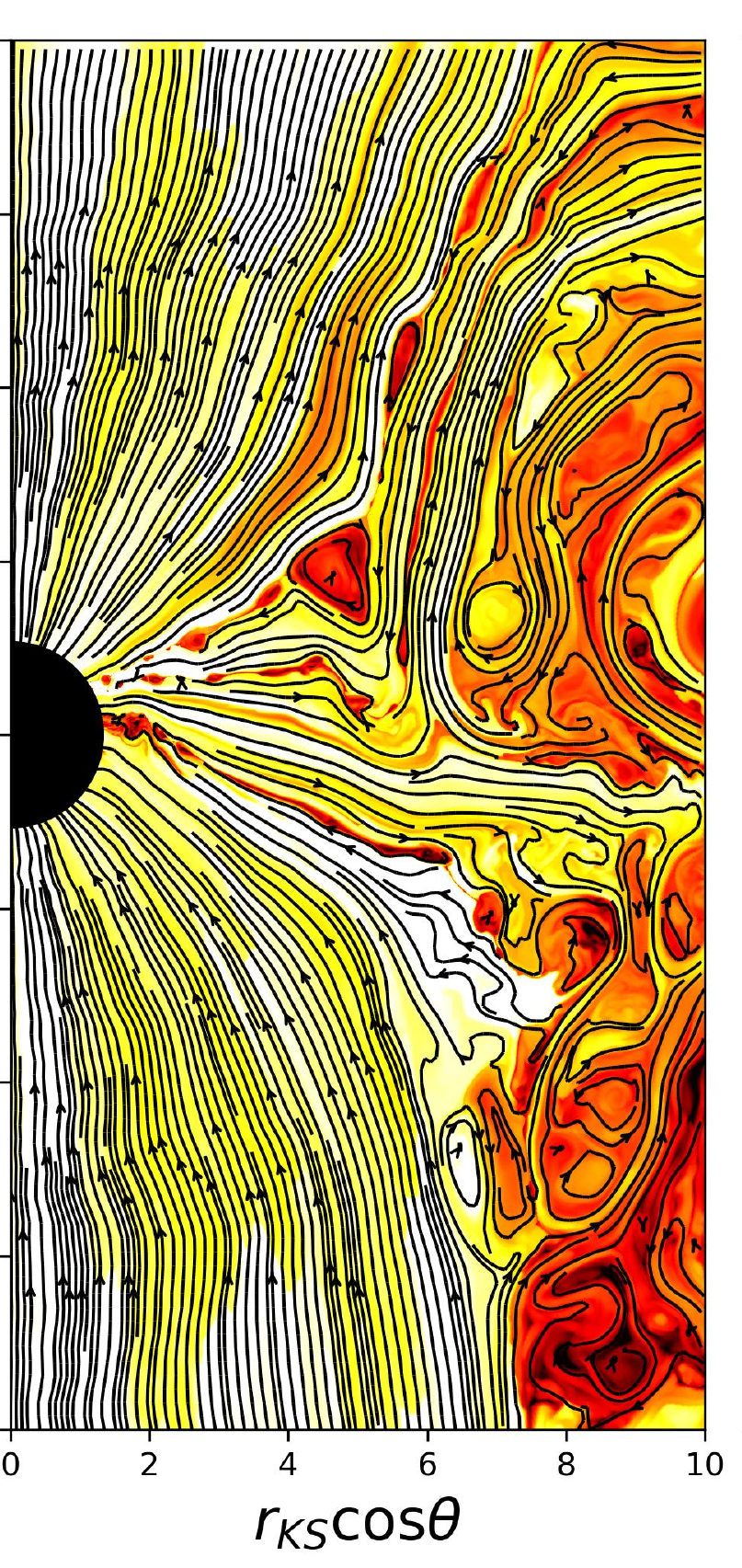}
    \includegraphics[width=0.2167\textwidth,trim= 15.2cm 1cm 14.7cm 2cm, clip=true]{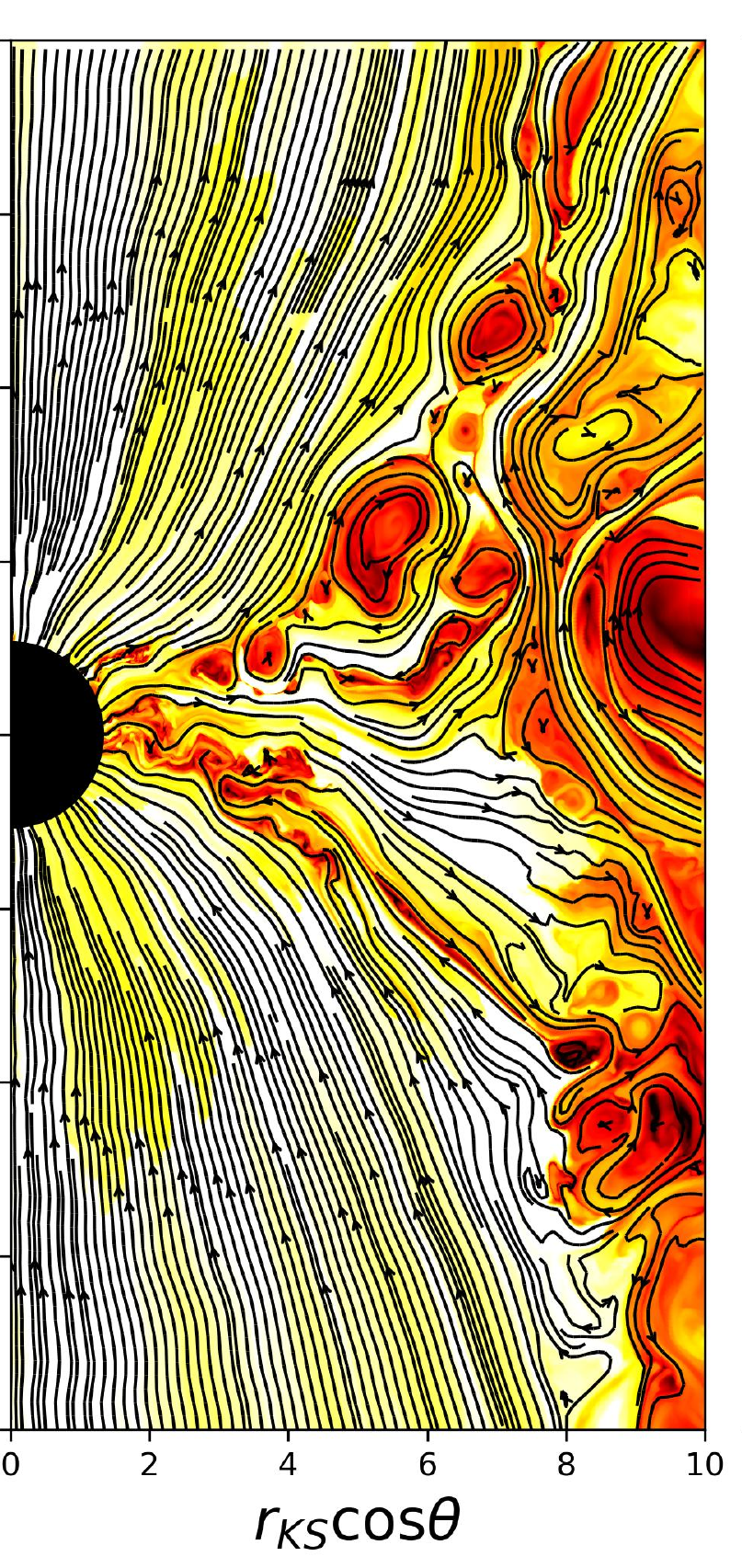}
    \includegraphics[width=0.275\textwidth,trim= 15.2cm 1cm 12.5cm 2cm, clip=true]{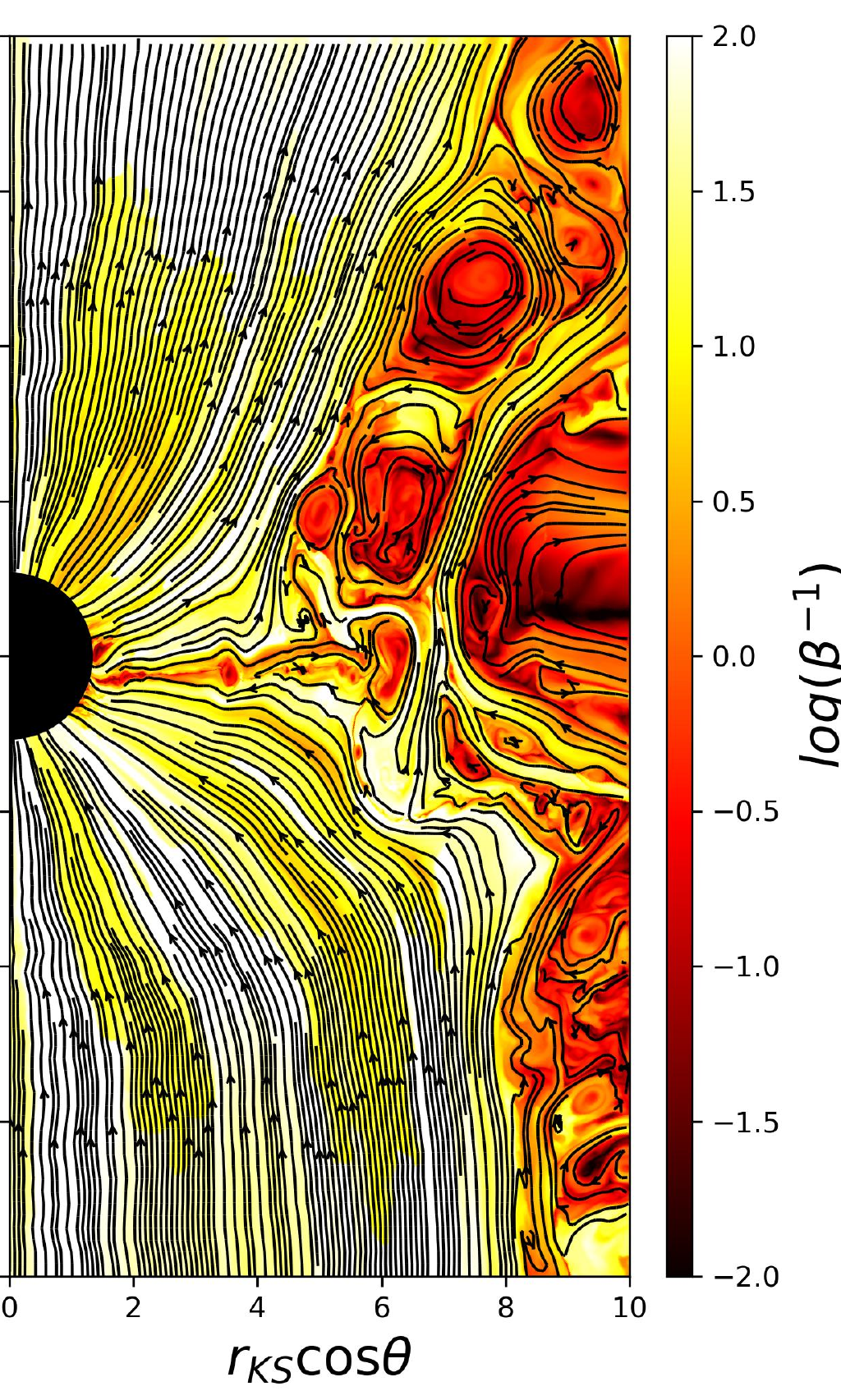}
    \caption{$\beta^{-1}=b^2/(2p)$ at four typical times $t = [2941, 2971, 2988, 3009] r_g/c$ (from left to right) during the quasi-state-state phase of accretion in the MAD configuration. Magnetic field lines are plotted on top as solid black lines. In the top half one can detect the accretion of a magnetic flux tube (left panel) at $x\approx3r_g,y\approx1r_g$ that opens up and becomes tearing unstable (second panel) after it connects to the black hole, and produces copious plasmoids coalescing into large-scale structures (third and fourth panels at $x\approx5r_g,y\approx2.5r_g$ with a typical size of about one Schwarzschild radius.}
    \label{fig:MAD287betaevolution}
\end{figure*}

In Figure \ref{fig:SANE154} we show the magnetization $\sigma = b^2/(\rho c^2)$, temperature $T = p/\rho$, magnetic dissipation $|E^i J_i|$, and the current density magnitude $J = \sqrt{J^i J_i}$, where $E^i$ and $J^i$ are the Eulerian electric field and current density (see \citealt{ripperda2019b} for definitions) at $t=1540 r_g/c$ (the third panel in Figure \ref{fig:SANE154betaevolution}). We mask the regions with $\sigma \geq 5$ where the flow dynamics is dominated by the density and pressure floors.
In the top-right panel one can detect ubiquitous current sheets along the jet's sheath and inside the disk within $10 r_g$, indicated by a strong current density. The current sheets are tearing-unstable and plasmoids form close to the event horizon after which they either fall into the black hole, or are advected along the jet's sheath, merge with other plasmoids, and grow into larger objects with a size of the order of a few $r_g$. 
The magnetization in the top-left panel shows that current sheets along the jet's sheath reconnect in a highly relativistic regime (i.e., $\sigma \gg 1$), whereas in the disk reconnection occurs in the transrelativistic regime $\sigma \lesssim 1$. 
Reconnection in the highly magnetized plasma surrounding the jet's sheath and the current sheets in the equatorial plane efficiently heats plasmoids to relativistic temperatures $T=p/\rho \sim 1$ as can be seen in the bottom-right panel. These hot plasmoids are advected along the jet's sheath or into the accretion disk. Heating of plasmoids that form inside the disk due to reconnection induced by MRI turbulence is less efficient due to the low magnetization in the disk. Instead, hot plasmoids observed at $x\approx6r_g,y\approx\pm2r_g$ were energized in the inner region close to the event horizon and ejected into the disk.
The plasmoids are mainly heated close to the event horizon by Ohmic heating as can be seen from the bottom-left panel. The strong parallel electric field indicated by $|E^i J_i|$ in the bottom-left panel can potentially accelerate particles to non-thermal energies.

Figure \ref{fig:SANE162} shows a zoom into the current sheet along the jet's sheath with several interacting plasmoids as visible Figure \ref{fig:SANE154}. The current sheets are captured by typically 10 cells along their widths, as can be seen from the on-plotted grid-block structure (white rectangles), where each block consists of $64\times32$ cells.

\subsection{Plasmoid formation in the MAD model}
In a MAD configuration the accretion disk is threaded by a strong magnetic flux suppressing the MRI and forming a more powerful magnetized jet than in the SANE case.
In Figure \ref{fig:MAD287betaevolution} we show the typical evolution of a current sheet indicated by a small $\beta^{-1}$ in the highly magnetized equatorial region close to the black hole at representative times $t = [2941,2971,2988,3009] r_g/c$ (from left to right). In the left panel a magnetic flux tube falls onto the black hole at $x \approx 1r_g, y\approx 3r_g$ and in the second panel the tube connects its two footpoints to the black hole. A near-equatorial current sheet forms and produces multiple small plasmoids within $5 r_g$ from the event horizon. The plasmoids that escape the gravitational pull merge into a large structure in the third panel at $x\approx 6r_g,y\approx 2.5r_g$. The large-scale plasmoid grows to a circular structure with a radius $\gtrsim 1 r_g$ and escapes along the jet's sheath at $x\approx 9r_g,y\approx 6r_g$ in the fourth panel. The process of formation to ejection takes place on a time scale of $\sim 70 r_g/c$. In the fourth panel the process restarts with an in-falling flux tube at $x\approx 9r_g, y\approx 0r_g$.

\begin{figure*}
    \centering
    \includegraphics[width=0.43\textwidth,trim= 13cm 1cm 12.5cm 2cm, clip=true]{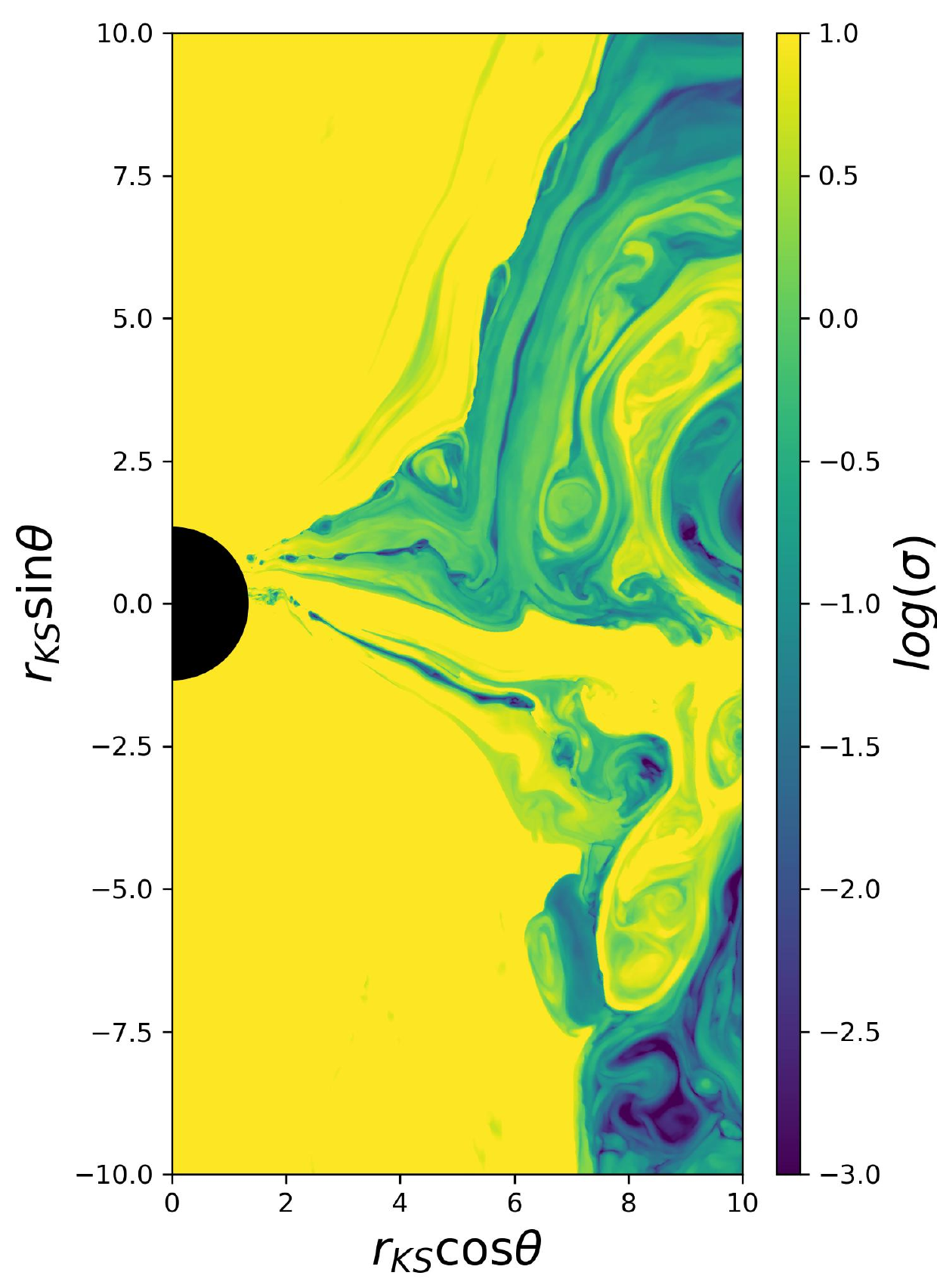}
    \includegraphics[width=0.43\textwidth,trim=13cm 1cm 12.5cm 2cm, clip=true]{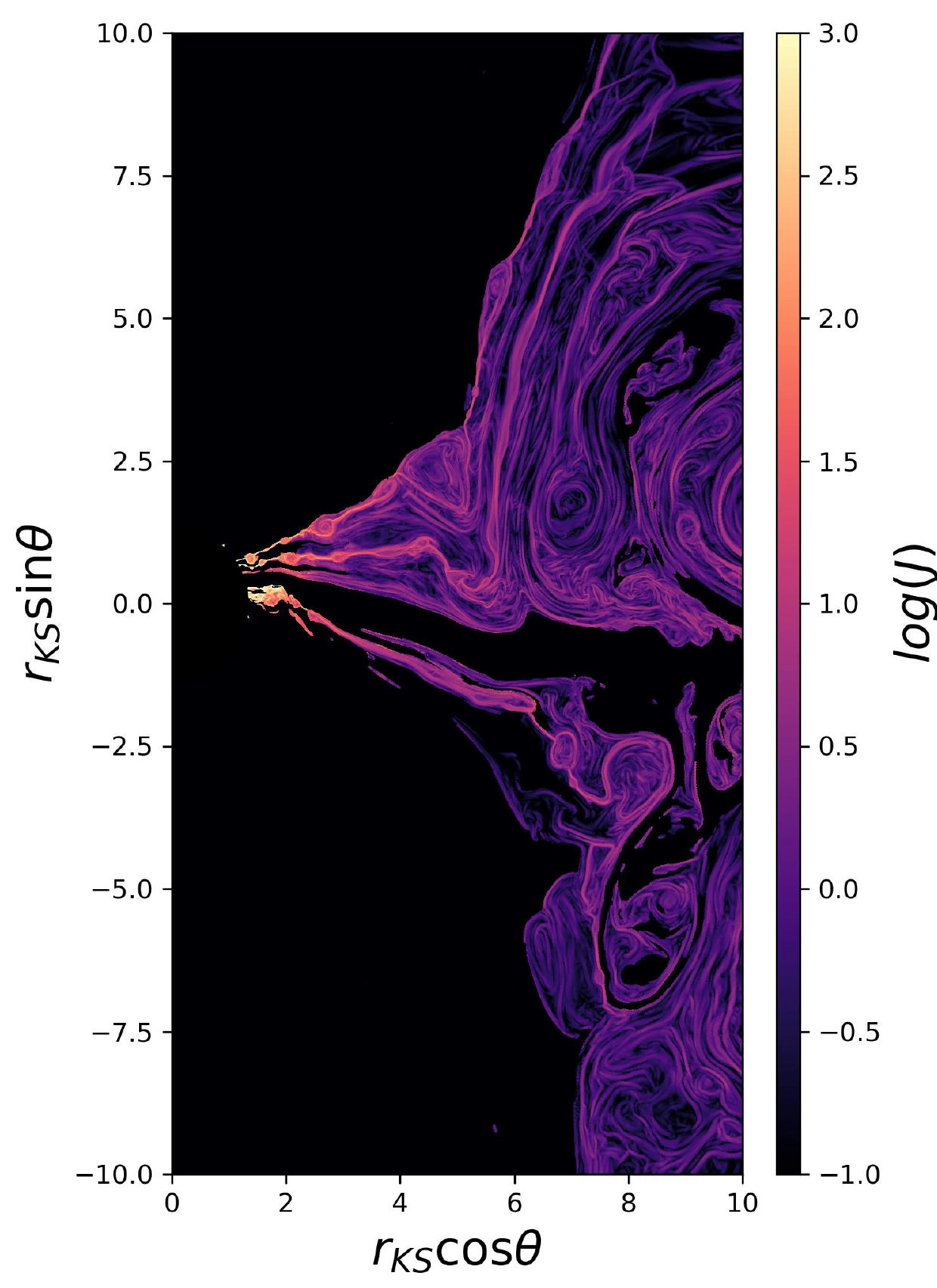}
    
    \includegraphics[width=0.43\textwidth,trim= 13cm 1cm 12.5cm 2cm, clip=true]{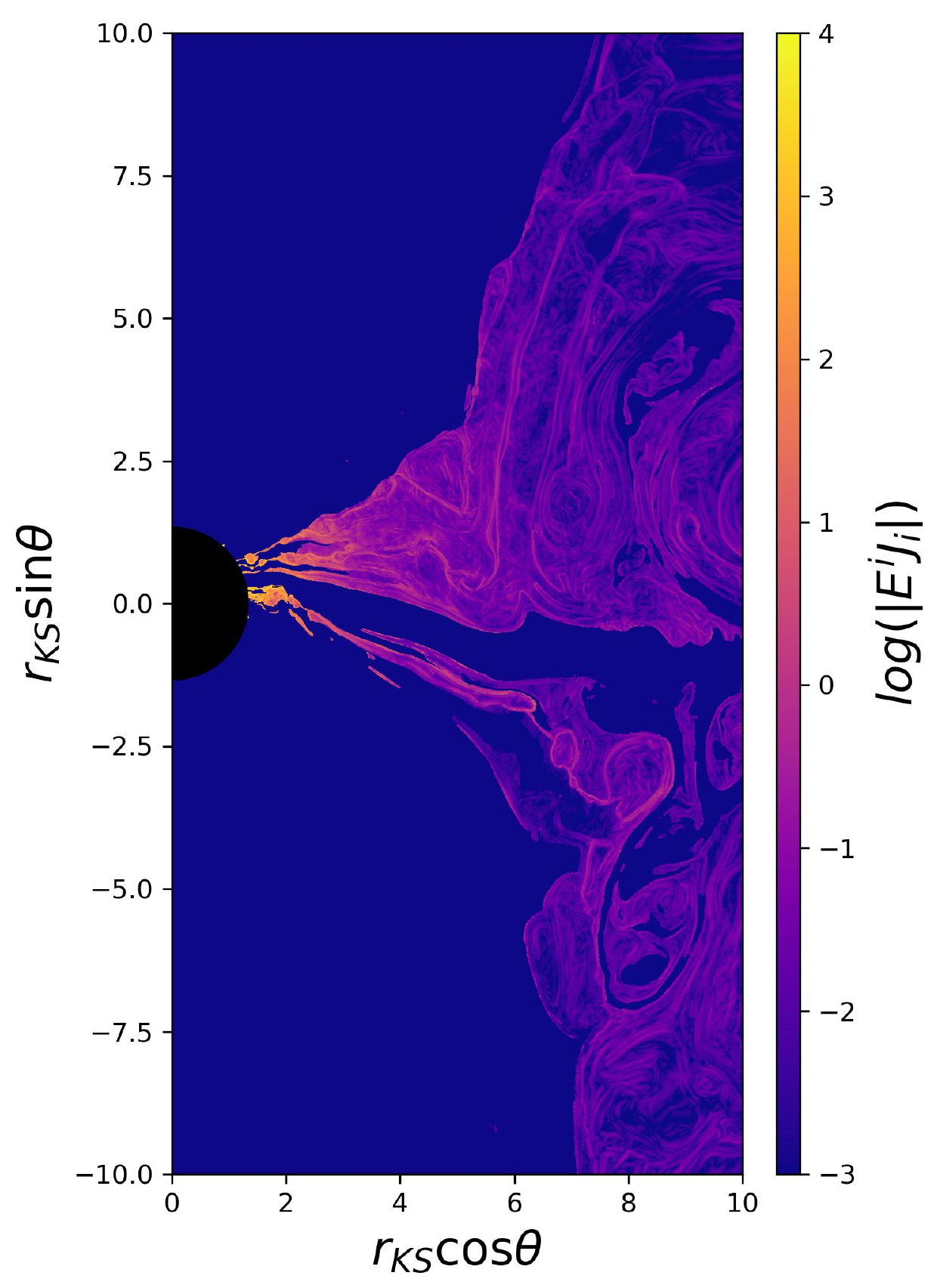}
    \includegraphics[width=0.43\textwidth,trim= 13cm 1cm 12.5cm 2cm, clip=true]{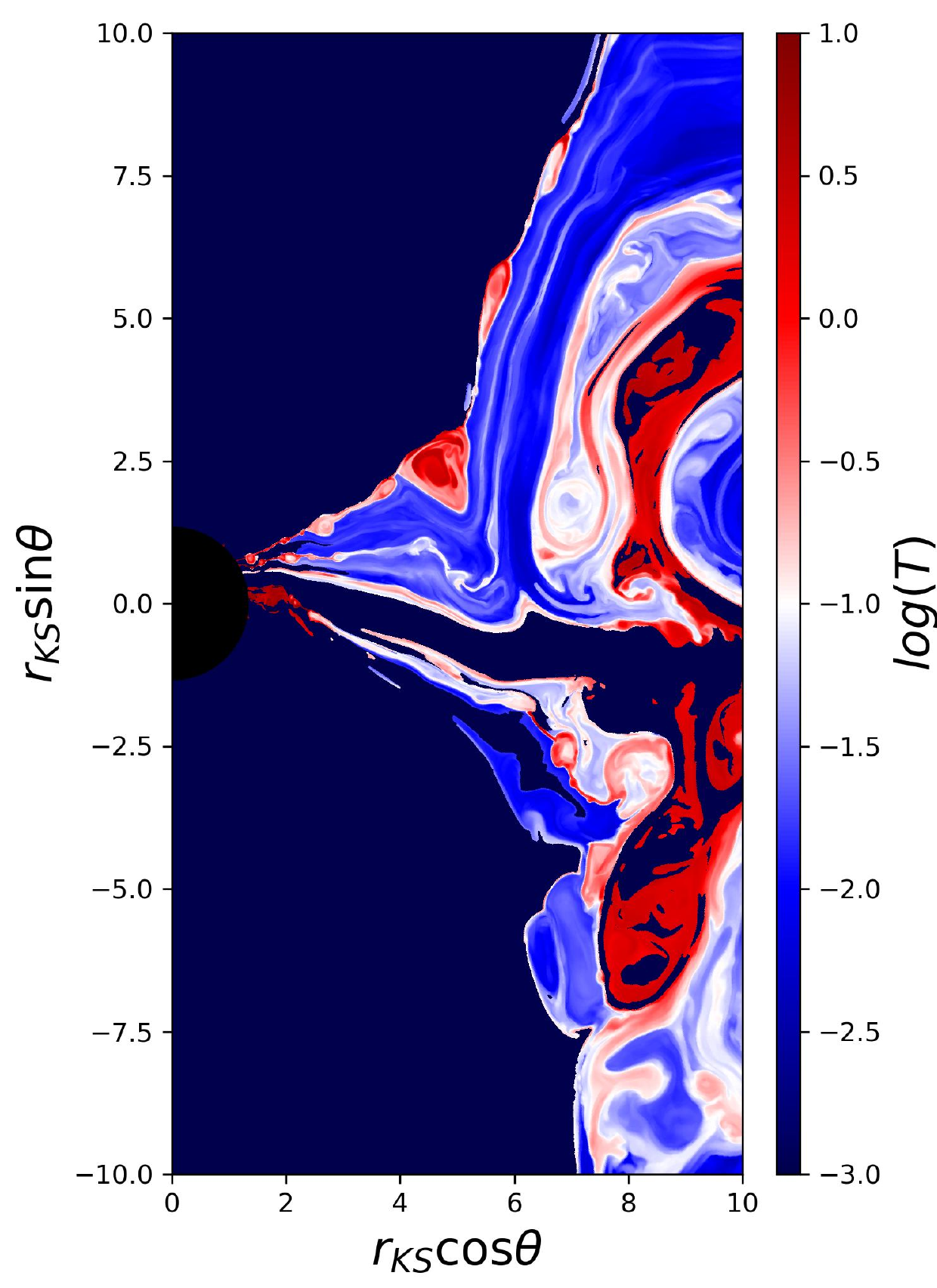}
    \caption{Representative MAD snapshot in the quasi-steady-state phase of accretion at $t = 2971 r_g/c$. top-left: Magnetization $\sigma$ showing that current sheets close to the event horizon are in the relativistic regime $\sigma \gg 1$.
    top-right: The thin tearing-unstable reconnection layers are indicated by a strong current density.
    bottom-left: Plasmoids in the current sheets are heated by Ohmic heating close to the event horizon.
    bottom-right: Plasmoids formed in the equatorial sheets are advected into both the disk and along the jet's sheath and are heated up to relativistic temperatures $T =p/\rho \sim 10$, an order of magnitude larger than in the SANE case in Figure \ref{fig:SANE154}.}
    \label{fig:MAD297}
\end{figure*}

In Figure \ref{fig:MAD297} we show the magnetization, current density, Ohmic heating, and temperature for the MAD state, where we again mask the regions with $\sigma \geq 5$.
Due to the higher magnetization (see the top-left panel) surrounding the equatorial current sheets (see the top-right panel) in the MAD state, the plasmoids are heated to relativistic temperatures $T \sim 10$ (bottom-right panel), an order of magnitude higher than in the SANE case. The current density in the sheets is significantly higher than in the SANE case. The plasmoids are heated through Ohmic heating close to the event horizon, as can be seen from the bottom-left panel.

\subsection{Reconnection rate}
\begin{figure}
    \centering
    \includegraphics[width=0.49\textwidth]{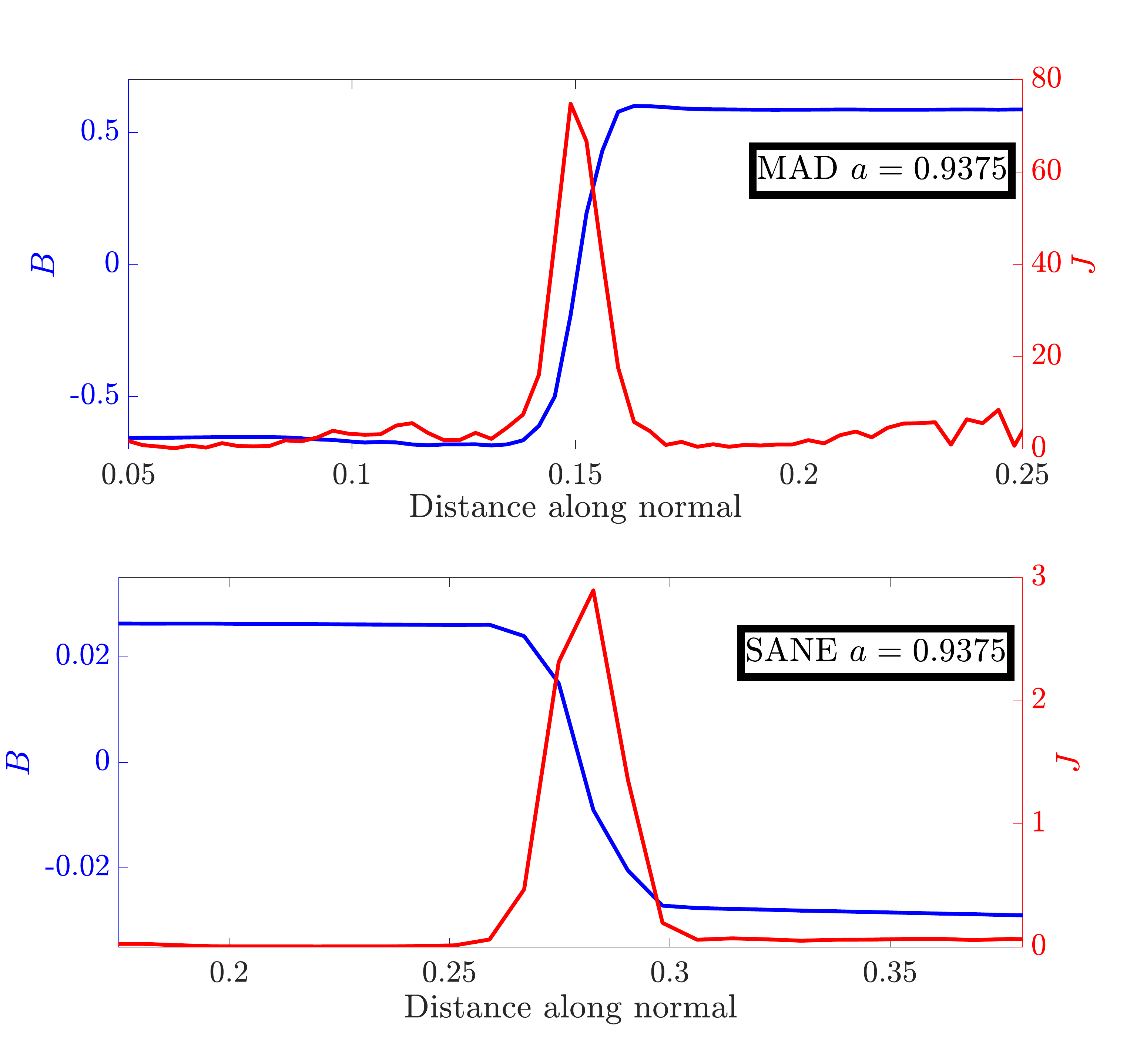}
    \caption{Magnetic field $B$ in a local inertial frame, projected along the sheet (blue line) and current density magnitude $J$ (red line) profiles taken on a slice through a current sheet in a MAD snapshot (top panel) and a SANE snapshot (bottom panel). The magnetic field geometry shows a typical current sheet profile.}
    \label{fig:MADSANEB2}
\end{figure}

We calculate the reconnection rate in a similar way as for the Orszag-Tang vortex for both MAD and SANE configurations. We first transform the Eulerian electric and magnetic fields into a local inertial frame (see e.g., \citealt{white2016}) to apply the standard reconnection analysis.
We project the fields in the flat frame along the direction parallel to the current layer to determine the upstream geometry, and a typical Harris-type sheet structure is found in Figure \ref{fig:MADSANEB2} both for the magnetic field and the current density magnitude $J$. All three magnetic field components switch sign in the current sheets, indicating that zero-guide-field reconnection occurs in both MAD and SANE cases.
In the local inertial frame we determine the inflow speed from the $\mathbf{E}\times\mathbf{B}$-velocity that we project along the direction perpendicular to the current sheet, and then calculate the reconnection rate as $v_{\rm rec}/c=(v_{\rm up, left}-v_{\rm up, right})/2c$. In both MAD and SANE configurations we select ten current sheets at different times during the quasi-steady-state phase of accretion and consistently find a reconnection rate between 0.01c and 0.03c. This finding is in accordance with analytic resistive MHD predictions for plasmoid-dominated reconnection in isolated current sheets (\citealt{bhattacharjee2009}; \citealt{uzdensky2010}). Note that the actual Lundquist number is approximately $S = L_{\rm{sheet}} c / \eta \gtrsim \mathcal{O}(10^5)$, since all current sheets have a typical length scale of $\sim \mathcal{O}(5-20 r_g)$, confirming that reconnection occurs in the plasmoid-dominated regime as $S \gg 10^4$.

\section{Flare analysis}
\begin{figure*}
    \centering
    \includegraphics[width=0.41\textwidth]{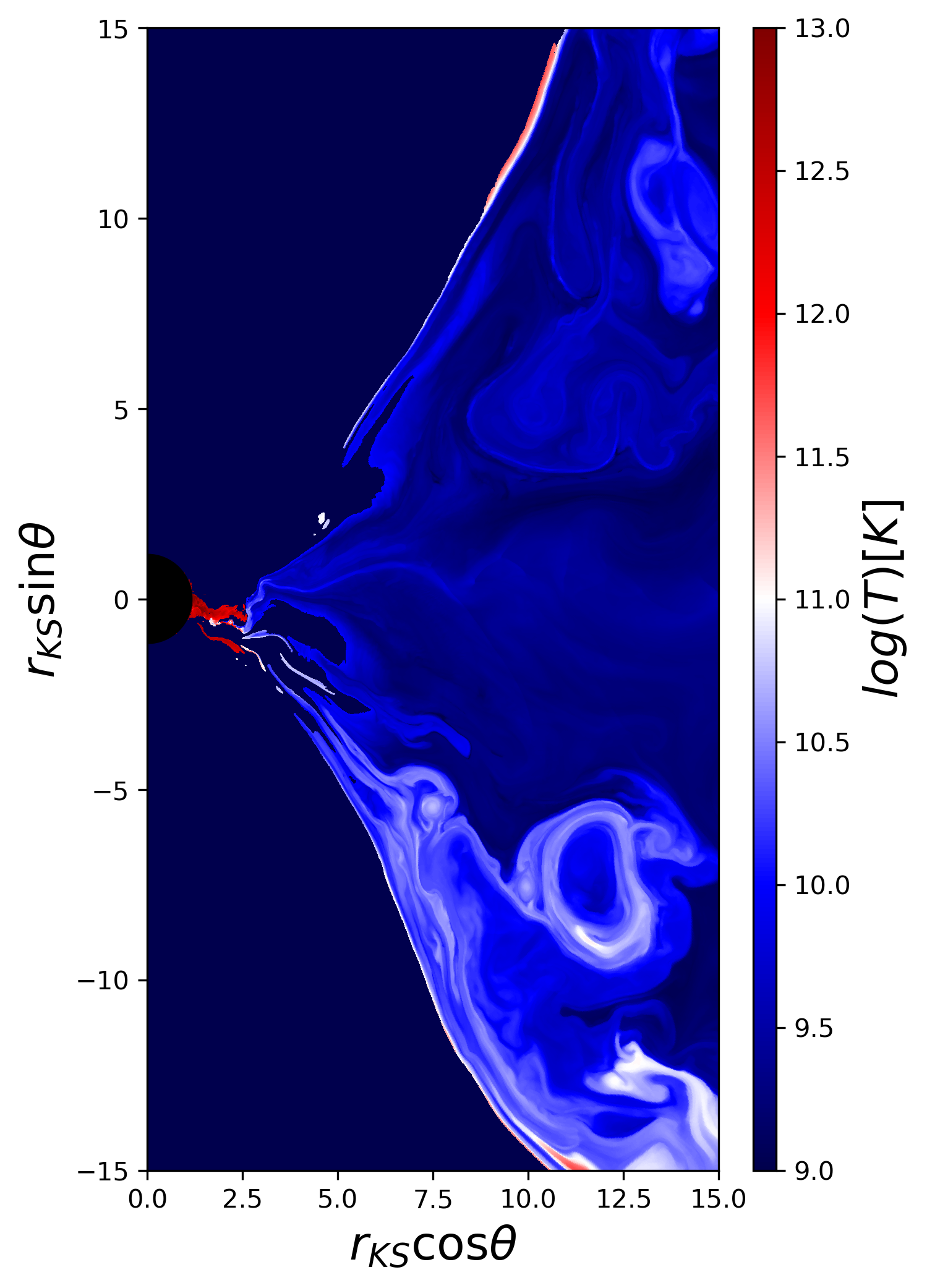}
    \includegraphics[width=0.41\textwidth]{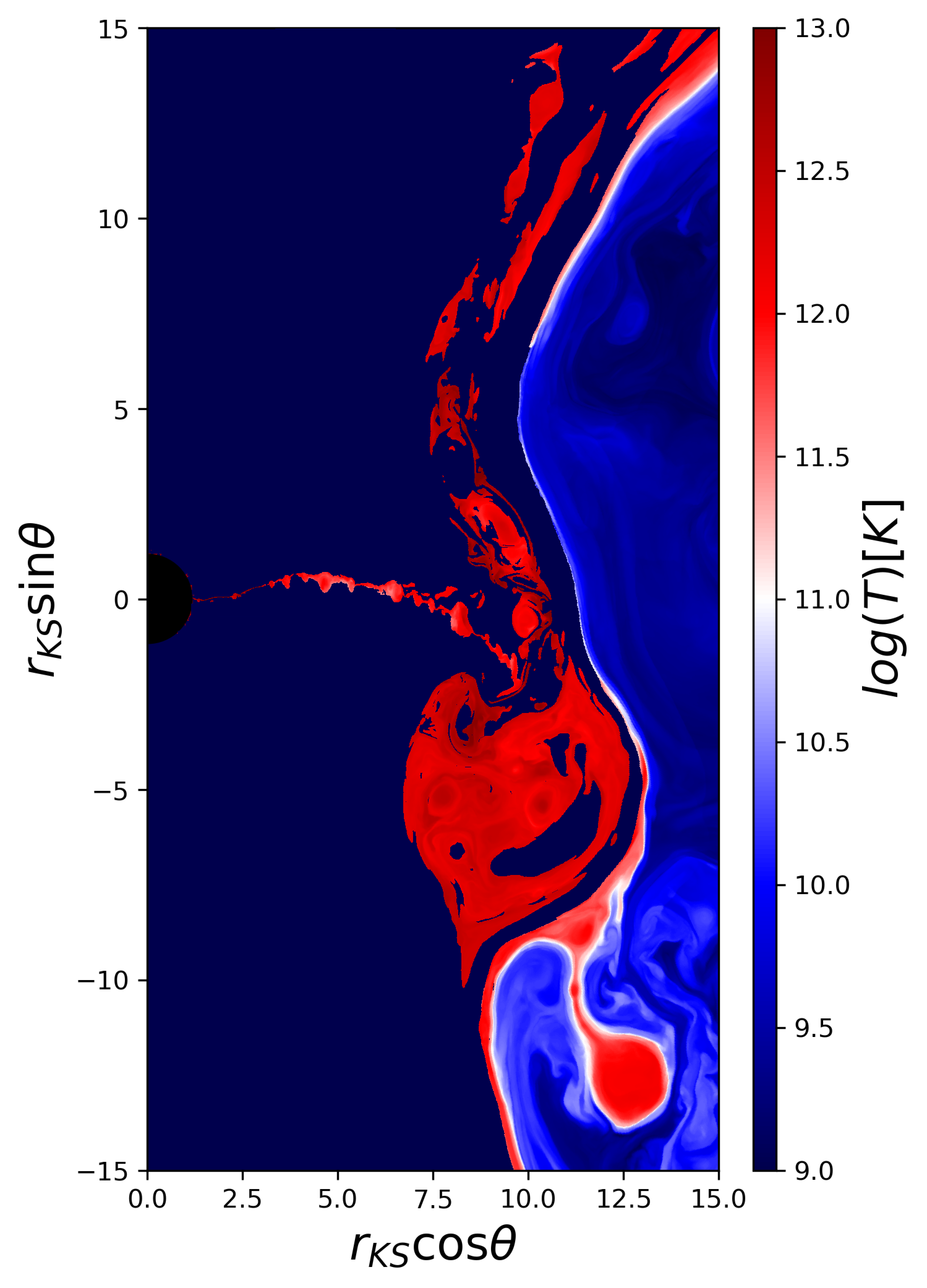}
    
    \includegraphics[width=0.41\textwidth]{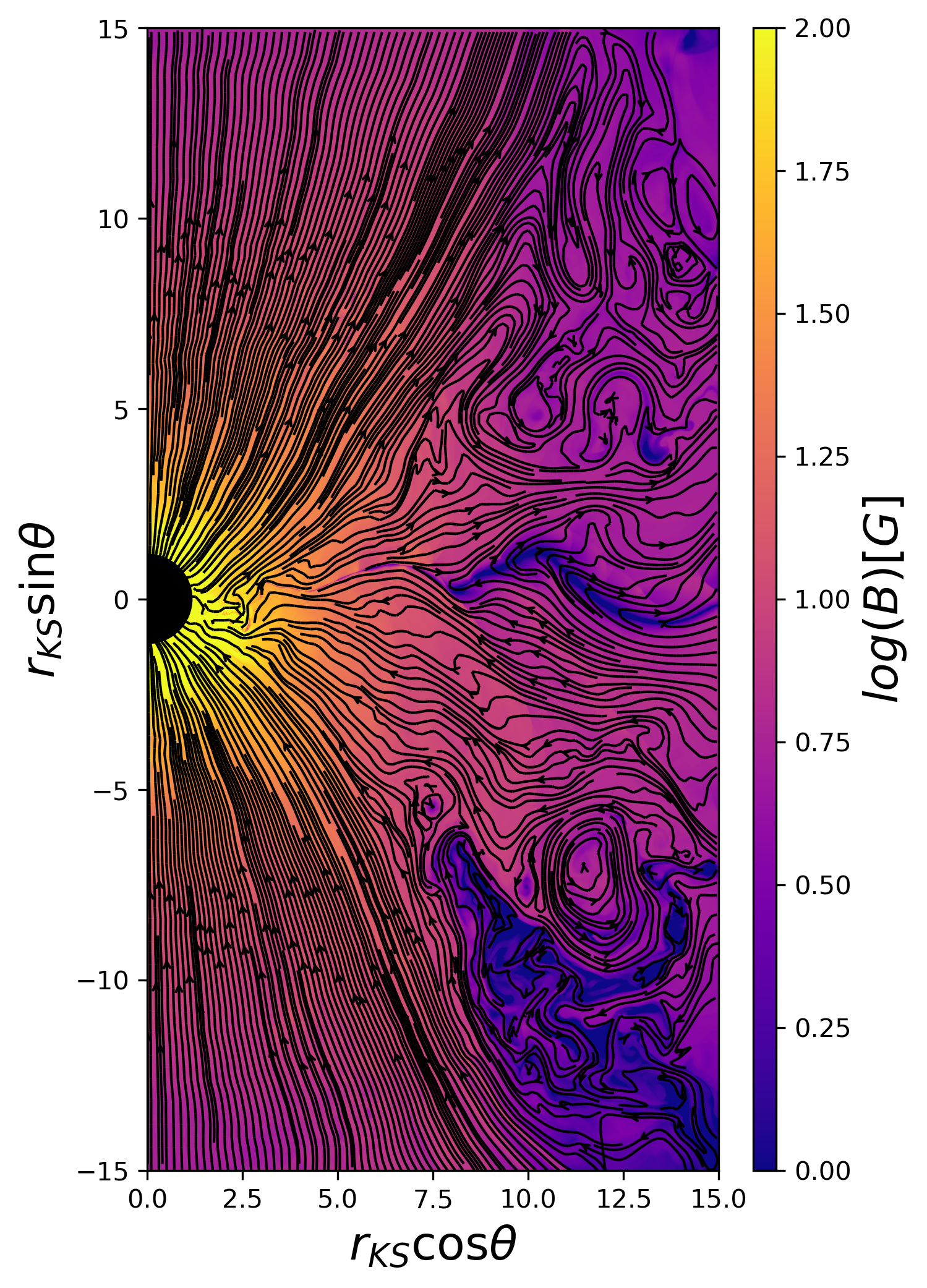}
    \includegraphics[width=0.41\textwidth]{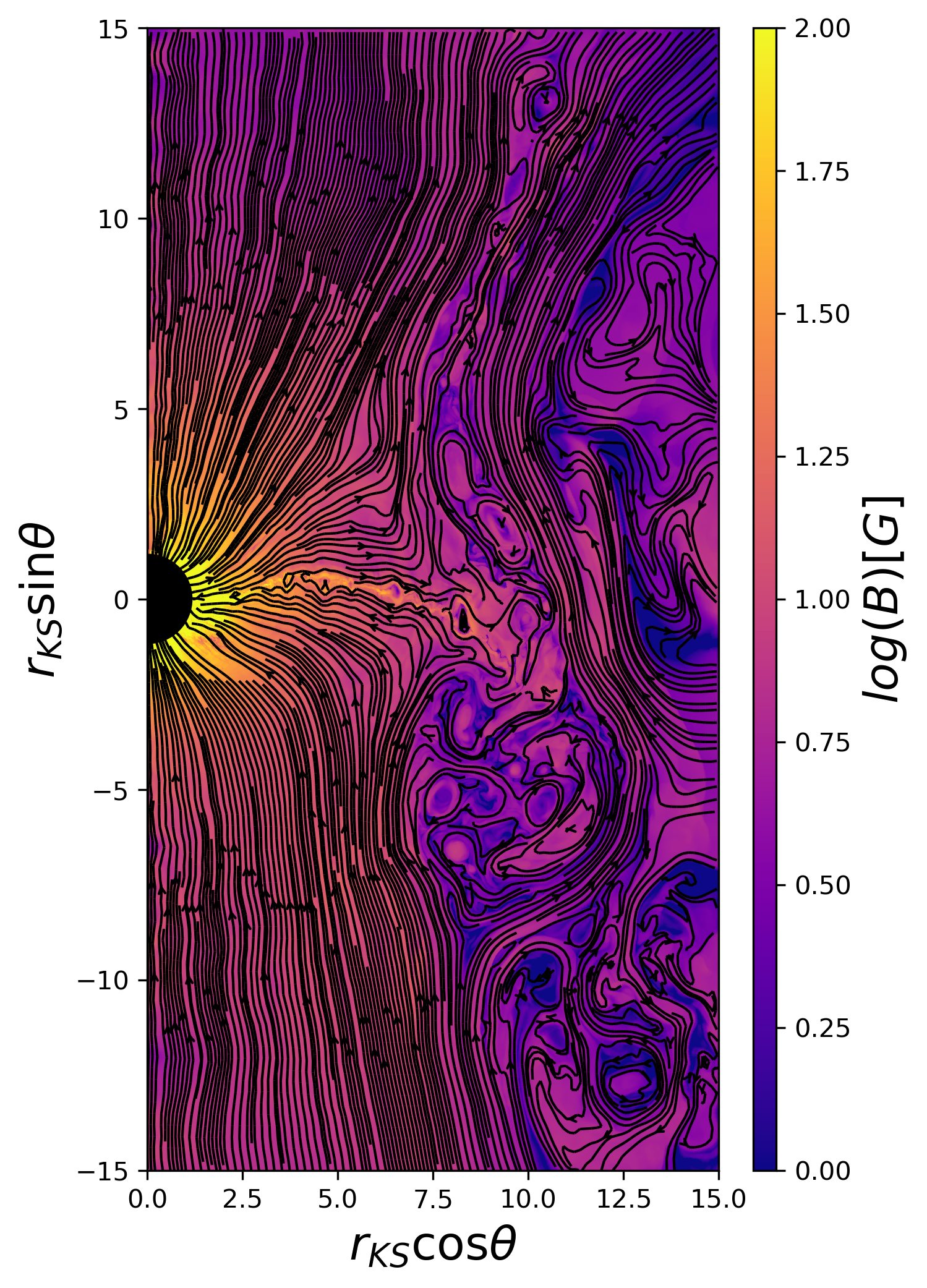}
    \caption{Representative prograde MAD snapshot for typical Sgr A$^*$ parameters during quiescence (left panels) at $t = 3155 r_g/c$ and during a flaring state (right panels) at $t = 3259 r_g/c$. top-left: The temperature $T$[K] in quiescent state remains approximately $10^{10}$K in the disk. A remnant of a previous flare (see Fig. \ref{fig:MAD297}) can be detected at $x\approx 12 r_g,y\approx-9 r_g$. top-right: A current sheet has formed at $y\approx 0 r_g$, producing plasmoids that heat the jet's sheath to temperatures $\gg 10^{10}$K through the reconnection exhaust. A hot spot with a radius of a few $r_g$ can be detected at $x\approx 8r_g,y\approx-5r_g$. bottom-left: The magnetic field during quiescence decreases as $\propto r^{-1}$ from approximately 50G at the inner radius of the disk at $\sim 3 r_g$ to 10G at $10 r_g$. bottom-right: The equatorial current sheet shows a clear field reversal. In the hot spot that is fed by the plasmoids from the current sheet, the magnetic field strength has decreased to 1-5[G], indicative of a large flare.}
    \label{fig:flare1}
\end{figure*}

Sgr A$^*$ shows daily flares in the near-infrared (on average every $\sim 6$ hours, \citealt{eckart2006}) and X-ray (on average every $\sim 12-24$ hours, \citealt{Baganoffflare03}) spectrum, often without a significant time-lag. The X-ray flares are large-amplitude outbursts followed by a quiescent period, whereas near-infrared flares appear as peaks within an underlying noise. The near-infrared flares can typically last for $\sim 80$ minutes, and the X-ray flares show shorter time-scales of $\sim 50$ minutes. Sub-structural variability with a characteristic timescale of $\sim 15–25$ minutes is regularly observed in near-infrared flares (\citealt{genzel2003,eckart2006}.
The \cite{Gravity2018} resolved the flare locations of three flares in the central $10 r_g$. The near-infrared flares are polarized, indicating their origin in synchrotron radiation produced by relativistic electrons. The polarization angle can change significantly during the flare (\citealt{Dodds_Eden2009,Doddseden2010}), indicating a change of topology of the magnetic field e.g., due to magnetic reconnection. The near-infrared flares in the spectrum are explained by a peak synchrotron frequency in accordance with Lorentz factors of $\Gamma \gtrsim 10^3(B/B_{\rm quiescence})^{-1/2}$, where $B$ is the magnetic field strength in Gauss [G] and $B_{\rm quiescence} \sim 10 - 50$G is the field strength in quiescent periods in the inner $10 r_g$ of the accretion disk (\citealt{Dodds_Eden2009}). This in turn requires particle acceleration, which is likely to be powered by tapping energy from the magnetic field, to energies well above the quiescent temperature of $\sim 3 \times 10^{10}$K (\citealt{Bower_2006}). The magnetic field strength is expected to significantly decrease to $\sim 1-10$G during a flare, to explain the simultaneity and symmetry of the X-ray and near-infrared light-curves (\citealt{Doddseden2010}).
\begin{figure}
    \centering
    \includegraphics[width=0.5\textwidth]{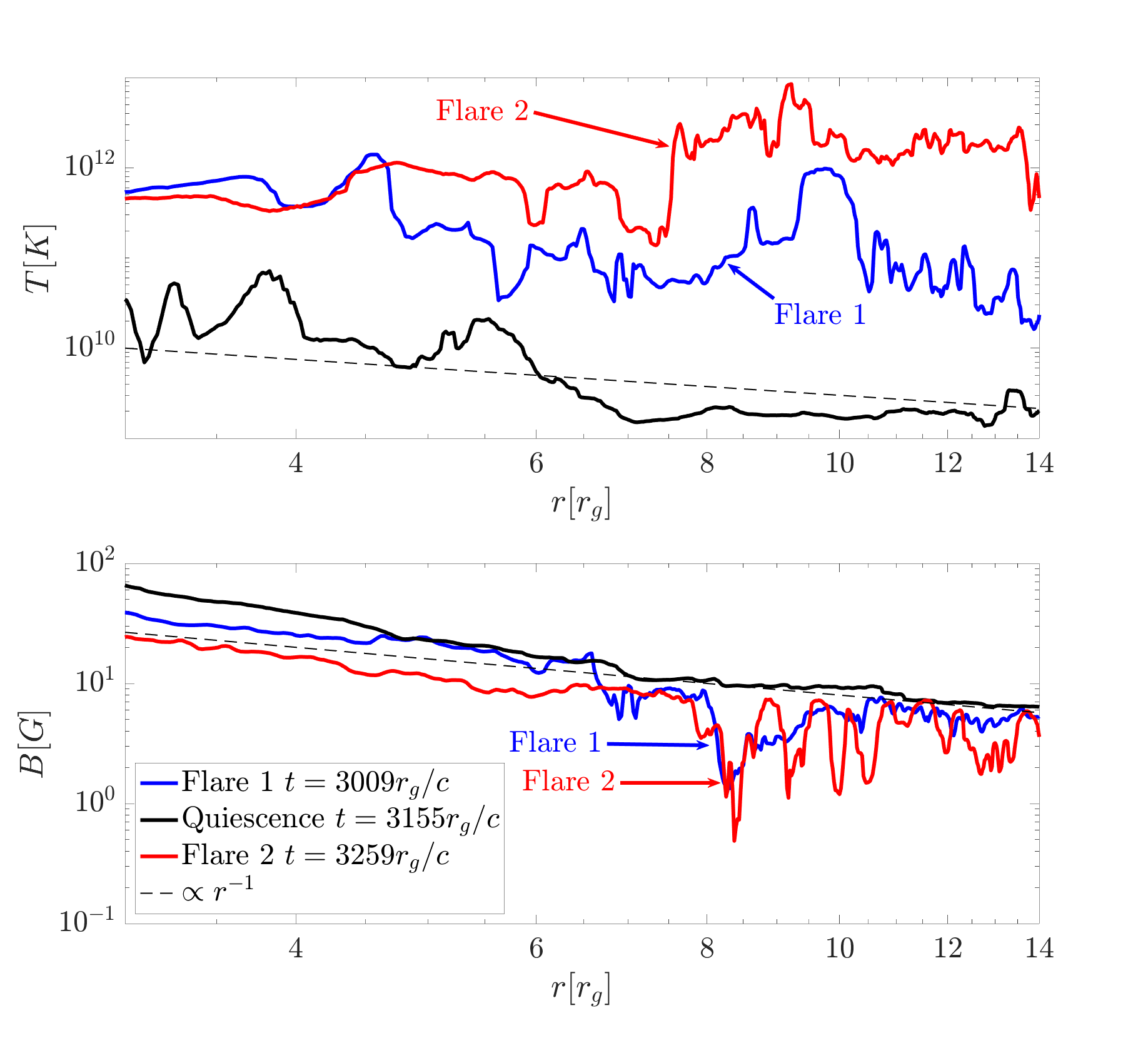}
    \caption{Radial cuts of the temperature $T$ and magnetic field strength $B$, assuming Sgr A$^*$ parameters, of the flares in Figs. \ref{fig:MAD287betaevolution} (flare 1, at $t=3009 r_g/c$) and \ref{fig:flare1} (flare 2, at $t=3259 r_c/c$). The black lines show $T$ and $B$ during quiescence. The magnetic field shows a clear drop from $\sim 10-20$G to $\sim 1-5$G between $7 r_g$ and $13 r_g$, while the temperature significantly increases, both for flare 1 and 2.}
    \label{fig:flare2} 
\end{figure}

Here, we compare the first time-dependent GRRMHD model for flare generation with observational constraints for Sgr A$^*$. To convert the plasma temperature, and magnetic field strength from code units to c.g.s.\ units we find a scaling factor $B_0 = 50$G for the MAD state such that the field strength at $3 r_g$, where the accretion disk starts, is equal to 50G in quiescence, as constrained by observations (\citealt{Dodds_Eden2009}) for Sgr A$^*$. This results in a field strength of 10G at $10 r_g$ in quiescence, and the field strength scales as $\propto r^{-1}$ with distance from the black hole. The fluid temperature $p/\rho$ is normalised as $T_0 = p_0 m_p / (\rho_0 k_B)$ [K], where $\rho_0 = B_0^2/(4\pi c^2 \sigma_0) \rm{[g/cm^3]}$, $p_0 = B_0^2 \beta_0 c^2 / (2 c^2 4\pi) {\rm [g/(cm \cdot s^2)]}$, $\sigma_0 = 0.33$, $\beta_0=0.35$ (taken at $3 r_g$), $m_p$ the proton mass, and $k_B$ Boltzmann's constant in c.g.s.\ units., resulting in a quiescent temperature of $\sim 10^{10}$K at $10 r_g$. Since the plasma is nearly collisionless, the electron temperature can be different from the temperature in the GRRMHD simulations, which has to be thought of as the proton temperature. For a normalization $\dot{M}_{0} = \rho_0r_g^2 c^3 / (MG)$ $\rm{[g/s]}$ we find a variable mass accretion rate of $\dot{M} \in [10^{-10},10^{-7}] M_{\odot} \rm{yr^{-1}}$ during the quasi-steady-state phase of accretion for both prograde and retrograde MAD simulations, consistent with bounds based on the measured Faraday rotation ruling out accretion rates greater than $\sim 10^{-7} M_{\odot} \rm{yr}^{-1}$ (\citealt{Bower2003,Marrone2007}). Note that we only determined a scaling factor $B_0$ for the magnetic field to find $\rho_0$, $p_0$, $T_0$ and $\dot{M}_0$, without relying on any other assumptions.

In Figure \ref{fig:flare1} we show the magnetic field $B$ [G] (bottom panels) and the temperature $T$ [K] (top panels) in quiescence (left panels) and during a flaring state (right panels). The temperature clearly increases by several orders of magnitude during a flare (top-right panel), and a hot spot forms at $x\approx 8r_g,y\approx-5r_g$. The hot spot is fed by plasmoids from the exhaust of the equatorial current sheet at $x\approx 8 r_g,y\approx0 r_g$. The plasmoids heat the jet's sheath, potentially explaining the mechanism behind limb-brightening which is for example observed down to $7r_g$ in the jet of the supermassive black hole in the center of the M87  galaxy (\citealt{Ly_2007,Kim_2018}).
The magnetic field clearly shows field reversal at the current sheet at $y\approx0 r_g$, and a decrease to $\sim 1-5$G in the hot spot. In quiescence the temperature remains of the order of $10^{10}$K in the inner $10 r_g$ (top-left panel), except in the inner $1 r_g$, where a new current sheet forms. A remnant of a previous flare (shown in \ref{fig:MAD297}) can be detected at $x\approx 12 r_g,y\approx-9 r_g$, and at $x\approx 14 r_g,y\approx10 r_g$ in the left panels.

In Fig. \ref{fig:flare2} we show the temperature (top panel) and the magnetic field strength (bottom panel) along a radial cut for two consecutive flares; The first one (the blue line) corresponds to the hot spot at $x\approx 7 r_g,y\approx6 r_g$ in Fig. \ref{fig:MAD287betaevolution} at $t=3009 r_g/c$; Then a quiescent period follows (corresponding to the black line, at $3155 r_g/c$); And then a second flare (the red line) at $x\approx 8r_g,y\approx-5r_g$, corresponding to the hot spot in Fig. \ref{fig:flare1}. Both flares originate from hot spots that have a radius of $\sim 1-3 r_g$, last for approximately $\sim 30-45$ minutes, and show clear substructure as a result from individual large plasmoids in the turbulent exhaust of the current sheet. The magnetic field drops within $\sim 10-15$ minutes from $10-50$G in quiescence to $\sim 1-5$G during a flare. This is consistent with a scenario where the synchrotron emission is produced by a non-thermal distribution of electrons, accelerated by energy that is tapped from the magnetic field via reconnection, following the analyses of \cite{Doddseden2010} and \cite{ponti2017} of multi-wavelength observations of the X-ray and near-infrared spectra of bright Sgr A$^*$ flares. The quiescence period of $\sim 60$ minutes is shorter than expected, which can potentially be explained by the axisymmetric nature of the simulations, causing an arrested inflow that is broken by bursts of accretion. In 3D simulations we expect the accretion to be more continuous, due to non-axisymmetric instabilities (\citealt{Igumenshchev2008,white2019}).

\cite{ponti2017} finds that $2\times 10^{35}$ erg $\rm{s^{-1}}$ is emitted in $30-60$ minutes during two observed simultaneous near-infrared and X-ray flares, resulting in a total energy release of $\gtrsim 10^{38}$ erg. We can estimate if the released magnetic energy in our prograde MAD simulation is roughly enough to power such a flare. We approximate the hot spot as a spherical emitting region where the magnetic field strength decreases (and temperature increases) from $B_{\rm quiescence} \approx 20$G to $B_{\rm flare} \approx 1$G within a radius of $r \sim 2 r_g$ for flare 1 and $r \sim 3 r_g$ for flare 2 (based on Fig. \ref{fig:flare2}). For Sgr  A$^*$, the Schwarzschild radius is $r_g \approx 6.1 \times 10^{11}$ cm. The total energy emitted is then equal to $\frac{4}{3} \pi r^3 (B_{\rm quiescence}^2-B_{\rm flare}^2)  / (8\pi) \gtrsim 10^{38}$ erg for both flares.

In our simulations of SANE accretion states plasmoids form constantly and interact with the turbulent disk, such that there is no clear distinction between a quiescent or a flaring state. For Schwarzschild black holes ($a=0$) plasmoids form in both MAD and SANE states, yet they never become powerful and large enough to heat the plasma significantly. \cite{Ressler_2020} recently showed that a MAD state is likely to form around Sgr A$^*$ by the continuous accretion being fed by 30 Wolf-Rayet stellar winds onto a central black hole.

\begin{figure}
    \centering
    \includegraphics[width=0.43\textwidth]{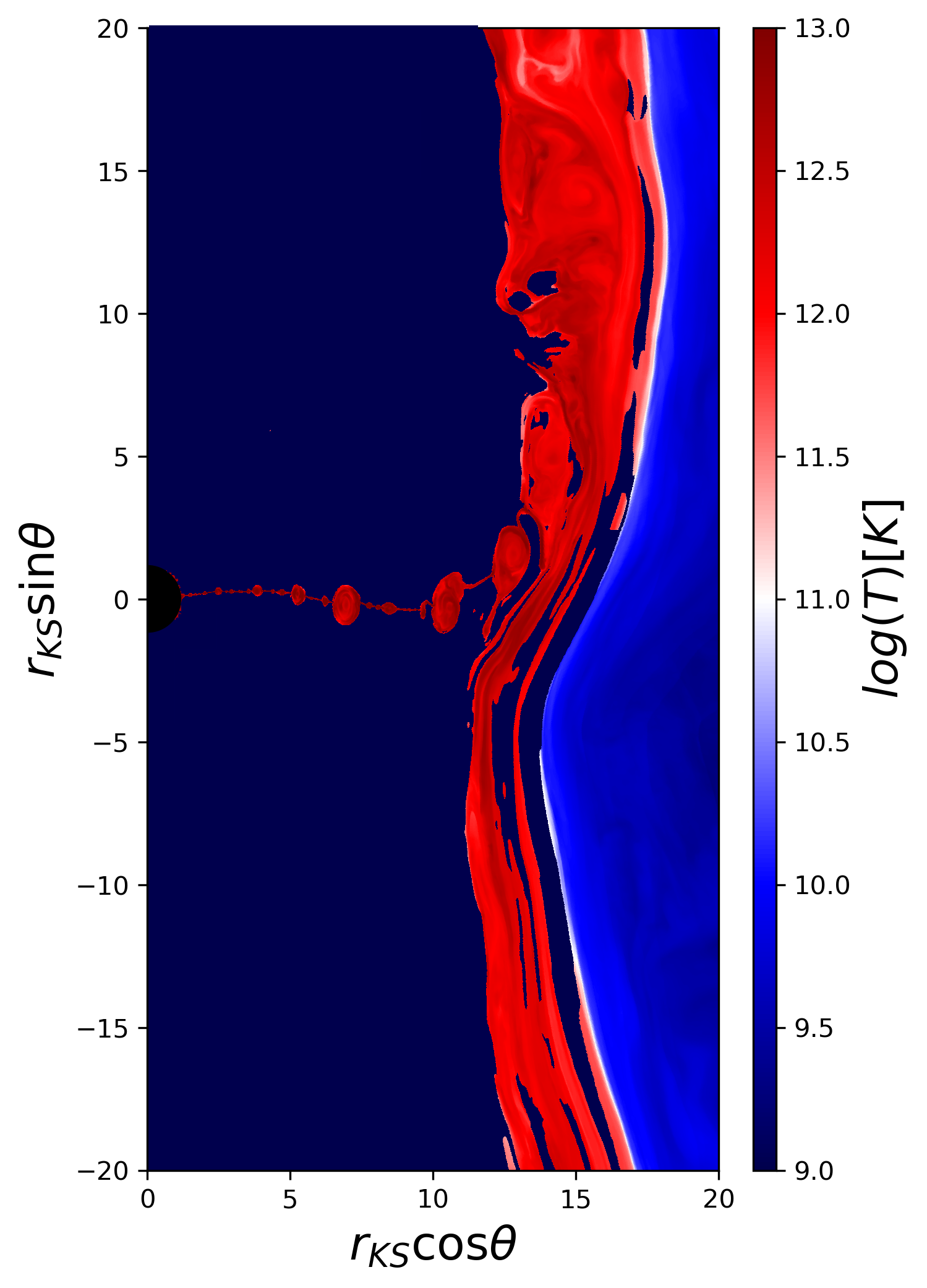}
    \caption{Limb-brightening of the jet sheath by hot plasmoids from the exhaust pf the equatorial current sheet. The color map shows the gas temperature, $T$[K], distribution at $t=3380 r_g/c$ in the MAD retrograde case with $a=-0.9375$.}
    \label{fig:MAD334_limb}
\end{figure}

In Fig. \ref{fig:MAD334_limb} we show the limb-brightened jet-sheat for a MAD accretion state with a retrograde spin $a=-0.9375$. The regions heated by plasmoids from the reconnection exhaust form further away compared to the prograde simulation\footnote{This can be related to the fact that the innermost stable circular orbit is further away for retrograde spins. The occurrence of hot spots only at $x \geq 15 r_g$ is potentially in conflict with the analysis of \cite{Gravity2018}.}.

\section{Discussion and conclusions}
We have shown that plasmoids form ubiquitously due to magnetic reconnection in black-hole accretion flows, regardless of the initial size of the disk and the magnetization during the quasi-steady-state phase of accretion. Energetic plasmoids that form on the smallest resistive scales and escape the gravitational pull of the black hole can grow into macroscopic hot spots through coalescence with other plasmoids. In both MAD and SANE cases these hot structures are ejected either along the jet's sheath or into the disk, heating the sheath and regions of the disk within a few Schwarzschild radii of the event horizon. In the MAD case the magnetization is significantly higher close to the event horizon, powering hot spots with relativistic temperatures $T=p/\rho \sim10$, an order of magnitude higher than in the SANE case $T\sim1$. 
The preferential heating of the jet's sheath by continuous reconnection events close to the event horizon can potentially explain Very Long Baseline Interferometry (VLBI) observations of the black hole at the center of the galaxy M87 which clearly show a limb-brightened jet down to $7r_g$ (\citealt{Ly_2007,Kim_2018}). 

In the SANE case current sheets and plasmoids also form in the turbulence induced by the MRI inside the disk. These plasmoids however do not reach relativistic temperatures due to the low magnetization in the disk. 

In both MAD and SANE cases circular hot spots can reach radii of approximately $1-3 r_g$ via plasmoid coalescence. The formation process from in-falling flux tubes to macroscopic hot spots of size $\sim \mathcal{O} (r_g)$ occurs on a time scale of $\sim \mathcal{O} (100 r_g/c)$, indicating a reconnection rate of $\sim 0.01c$. By analyzing multiple individual current sheets, we confirm reconnection rates between $0.01c$ and $0.03c$ for Lundquist numbers of $S \gtrsim 10^5$, in accordance with numerical relativistic resistive MHD studies of isolated plasmoid-dominated reconnection (e.g.,\ \citealt{delzanna2016,ripperda2019}) and analytic predictions (\citealt{bhattacharjee2009}; \citealt{uzdensky2010}).

In all our simulations, regardless of spin and accretion state, plasmoids with a size of $\sim 1 r_g$ form within $5-10 r_g$ from the event horizon on time scales of $100 r_g/c \approx 30$ minutes assuming conditions of Sgr A$^*$, which is in accordance with orbiting hot spots observed by \cite{Gravity2018,gravity2020}. Although in reality hot spots orbit around the black hole, in our simulations we assume axisymmetry such that analyzing trajectories in the invariant azimuthal ($\phi$-)direction cannot be directly related to observations. 
Only MAD accretion disks around spinning black holes are consistent with observations of near-infrared and X-ray flares, where large magnetic field amplitudes of $10-50 $G are observed in quiescence, and the field strength drops to $1-5$G at the peak of a flare (\citealt{Dodds_Eden2009,Doddseden2010,ponti2017}), 
due to conversion of the magnetic energy into accelerated electrons and synchrotron emission through reconnection. We find that during a flare period in our MAD simulations energies $\gtrsim 10^{38}$ erg can be emitted from a spherical region of $1-3 r_g$ within $30-45$ minutes. The flares are associated with events of reduced magnetic flux through the horizon and periodic reformation of the accretion flow (see the top panel of Fig. \ref{fig:mdot}). Hints of the same process of cycles of flux build-up and dissipation have been observed in 3D MAD simulations by \cite{dexter2020sgr} and \cite{porth2020flares}.
In our high-resolution SANE simulations there is no substantial variability in the current sheet and plasmoid formation which could be robustly associated with a flaring event.

Assuming axisymmetry prevents us from resolving non-axisymmetric instabilities that can potentially disrupt the current sheets forming close to the event horizon. In future work we plan to address whether plasmoid-structures can form in full 3D simulations and if they can result in orbiting hot spots. In 3D simulations we plan to trace test particles to model radiative signatures with realistic electron distribution functions (\citealt{ripperda2017,ripperda2019c,bacchini2018,bacchini2019}). The radiative properties of Sgr A$^*$ have been studied in ideal GRMHD simulations by \cite{Davelaar_2018}, concluding that 5-10\% of the electrons inside the jet's sheath are expected to be accelerated to non-thermal distributions during a flare.

In this work we rely on numerical viscosity, such that the typical magnetic Prandtl number, indicating the ratio between viscosity and resistivity, is smaller than unity. In a future study we will analyze the effect of adding explicit viscosity and resistivity on plasmoid formation in MRI turbulence in shearing box simulations considering a range of Prandtl numbers. 

\section*{Acknowledgements}
The computational resources and services used in this work were provided by facilities supported by the Scientific Computing Core at the Flatiron Institute, a division of the Simons Foundation; And by the VSC (Flemish Supercomputer Center), funded by the Research Foundation Flanders (FWO) and the Flemish Government – department EWI. 
BR is supported by a Joint Princeton/Flatiron Postdoctoral Fellowship. FB is supported by a Junior PostDoctoral Fellowship (grant number 12ZW220N) from Research Foundation -- Flanders (FWO). Research at the Flatiron Institute is supported by the Simons Foundation. We would like to thank Amitava Bhattacharjee, Luca Comisso, Jordy Davelaar, Charles Gammie, Yuri Levin, Matthew Liska, Nuno Loureiro, Sera Markoff, Koushik Chatterjee, Elias Most, Kyle Parfrey, Lorenzo Sironi, Amiel Sternberg, Jim Stone, and Yajie Yuan for useful discussions.
\bibliography{mylib3}{}
\bibliographystyle{aasjournal}

\end{document}